\documentclass[10pt,twocolumn,twoside,journal]{IEEEtran}
\IEEEoverridecommandlockouts
\usepackage{subfigure} 
\usepackage{graphicx}

\usepackage{amsmath,graphicx,amssymb,mathtools,bm}
\usepackage{subfigure}
\usepackage{hyperref}
\usepackage{cite}
\usepackage{amsmath,amssymb,amsfonts}  
\usepackage{textcomp}
\usepackage{xcolor}
\usepackage{verbatim}  
\usepackage{bm}  
\usepackage{mathrsfs} 
 \usepackage{algorithmic} 
\usepackage{booktabs}
\usepackage{textcomp}  
\usepackage{multirow}  
\usepackage{lettrine}   
\usepackage{graphicx}  
\usepackage{color}  
\usepackage{amsmath}
\usepackage{amssymb}
\usepackage{stfloats} 
\usepackage{caption} 

\usepackage{color}  
 \usepackage[ruled,linesnumbered]{algorithm2e}

\begin{document}
	\newcommand{\tabincell}[2]{\begin{tabular}{@{}#1@{}}#2\end{tabular}}
   \newtheorem{Property}{\it Property} 
  
 \newtheorem{Proposition}{\bf Proposition}
\newtheorem{remark}{Remark}
\newenvironment{Proof}{{\indent \it Proof:}}{\hfill $\blacksquare$\par}

\title{Flexible WMMSE Beamforming for MU-MIMO\\ Movable Antenna Communications 
}
 
\author{Songjie Yang, Zihang Wan, Yue Xiu, Boyu Ning, Yong Li,\\
	Yuenwei Liu,   \IEEEmembership{Fellow,~IEEE}, Chau Yuen,  \IEEEmembership{Fellow,~IEEE}


\thanks{
	Songjie Yang, Yue Xiu, Boyu Ning are with the National Key Laboratory of Wireless Communications, University of Electronic Science and Technology of China, Chengdu 611731, China. (e-mail:	yangsongjie@std.uestc.edu.cn; xiuyue12345678@163.com; boydning@outlook.com;).
	
	Zihang Wan is with the Key Laboratory of Wireless Optical Communication, Chinese Academy of Sciences, School of Information Science and Technology, University of Science and Technology of China, Hefei 230026, China (e-mail: zihangwan@mail.ustc.edu.cn)
	
	Yong Li is with the Henan Agricultural Environment Monitoring and Big Data Analysis Engineering Research Center, Henan Agricultural University, Zhengzhou 450046, China. (e-mail: ly@henau.edu.cn).
	
	Yuanwei Liu is with the Department of Electrical and Electronic Engineering, The
	University of Hong Kong, Hong Kong (email: yuanwei@hku.hk).
	
		Chau Yuen is with the School of Electrical and Electronics Engineering, Nanyang Technological University (e-mail: chau.yuen@ntu.edu.sg).}}

\maketitle

\begin{abstract}
Movable antennas offer new potential for wireless communication by introducing degrees of freedom in antenna positioning, which has recently been explored for improving sum rates. In this paper, we aim to fully leverage the capabilities of movable antennas (MAs) by assuming that both the transmitter and receiver can optimize their antenna positions in multi-user multiple-input multiple-output (MU-MIMO) communications. Recognizing that WMMSE beamforming is a highly effective method for maximizing the MU-MIMO sum rate, we modify it to integrate antenna position optimization for MA systems, which we refer to as flexible WMMSE (F-WMMSE) beamforming. Importantly, we reformulate the subproblems within WMMSE to develop regularized sparse optimization frameworks to achieve joint beamforming (antenna coefficient optimization) and element movement (antenna position optimization). We then propose a regularized least squares-based simultaneous orthogonal matching pursuit (RLS-SOMP) algorithm to address the resulting sparse optimization problem. To enhance practical applications, the low-complexity implementation of the proposed framework is developed based on the pre-calculations and matrix inverse lemma.  The overall F-WMMSE algorithm converges similarly to WMMSE, and our findings indicate that F-WMMSE achieves a significant sum rate improvement compared to traditional WMMSE, exceeding $\textbf{20\%}$ under appropriate simulation conditions.

\end{abstract}
\section{Introduction}

As wireless communication technology evolves, continuous advancements are being made across various dimensions to enhance degrees of freedom (DoFs) and expand applications. A key focus is the exploration of wireless channel characteristics, leveraging their properties to improve communication and sensing capabilities. This includes utilizing the sparsity of millimeter-wave and terahertz channels for beamspace signal processing \cite{mmw1, mmw2, BN1}, employing reconfigurable intelligent surfaces (RIS) to enhance channel propagation \cite{RIS1, RIS2, BN2}, and exploiting near-field spherical-wave channels to unlock the distance DoF \cite{NF1, NF2}. Despite these developments, the potential of wireless channels remains a vast area for further exploration.

With the emergence of shape-adjustable, controllable, and reconfigurable materials and antennas, flexible radio systems hold significant promise for the future. These innovations are primarily driven by researchers in the antenna domain and are being applied across various fields, such as wearable devices and dynamic deployment designs. Advances in flexible antennas---encompassing positional movement, shape control, rotational adjustment, and pattern reconfiguration---	are documented in \cite{Liquid, MEMS, Pixel, Stepper}. Recognizing the potential of flexible antenna DoFs, recent studies have investigated wireless communication with variable antenna positions. The primary focus has been on movable antennas (MAs) and fluid antennas (FAs), both of which optimize antenna element positions to leverage the inherent capabilities of wireless channels \cite{MA4, FA1}. In contrast to conventional fixed-position antennas, MAs and FAs can proactively reshape wireless channels into more favorable conditions for data transmission, avoiding positions or angles that may experience deep fading for desired users or strong interference for undesired ones. Typically, research on MAs centers on spatial geometry channels, while studies on FAs focus on statistical channels.

It is worth noting that before the advent of MAs and FAs, two dynamic antenna position optimization techniques were extensively explored in the field of signal processing. The first technique, antenna selection \cite{AS1}, involves identifying the optimal subset of antennas from a predefined dense array, effectively constituting discrete position optimization. Initially developed to maximize energy efficiency in wireless communication systems, this technique is also commonly applied in spatial index modulation, where the best subset of antennas is selected for transmission. Additionally, in integrated sensing and communication systems, antenna selection can optimize which antennas are designated for transmitting signals and which are used for receiving echo signals. From the perspective of array structure exploitation, antenna selection falls under discrete antenna position optimization schemes. The second technique, array synthesis \cite{ars1, ars2}, focuses on achieving specific beam pattern objectives, such as sidelobe suppression, main lobe directivity enhancement, and efficient beam pattern representation using fewer elements through the optimization of antenna positions, counts, and beamforming coefficients. Although differing in objectives and methodologies, the strategies employed in antenna selection and array synthesis offer valuable insights for optimizing flexible arrays.

Recently, several studies for MAs and FAs were investigated in \cite{FA1,FA2,FA3,MA1,MA2,MA3}. In \cite{MA1}, the authors maximized the multi-path channel gain under MA systems and showed the periodic behavior
of the multi-path channel gain in a given spatial field, providing insights for MA-enhanced communications.
The authors of \cite{FA1} derived a closed-form expression for the lower bound of capacity in FASs, confirming the substantial capacity gains derived from the diversity concealed within the compact space. Both \cite{MA1} and \cite{FA1} theoretically showcased the potential benefits of antenna position optimization in wireless communications. 
The study in \cite{FA4} explored point-to-point FAS communications using maximum ratio combining, revealing that the system's diversity order matches the total number of ports.
Moreover, \cite{FA2} noted in FASs that the multiplexing gain was directly proportional to the number of ports and inversely related to the signal-to-interference ratio target. Meanwhile, \cite{MA2,FA3} showed the potential of uplink power minimization through optimizing the user's antenna position. Conversely, \cite{MA3} explored MAs at the base station (BS) end, utilizing particle swarm optimization to optimize antenna positions with the objective of maximizing the minimum user rate. However, such meta-heuristic algorithms, despite their effectiveness, often come with high computational complexity and may lack deeper theoretical insights. For more insights, \cite{FPR} proposed flexible precoding with joint antenna coefficient and position optimization based on a sparse optimization framework, which provided a new view for MAs.
Moreover, flexible antenna DoF was also combined with other wireless applications to unlock new potentials, such as integrated sensing and communication (ISAC) \cite{MAISAC1,MAISAC2,MAISAC3} and mobile edge computing \cite{MAMEC}. In these references, MAs have been demonstrated useful for system performance enhancement.

Beyond element-movable cases reviewed above, some works have been focused on array-level adjustment to explore  array DoFs. More specifically, \cite{MAR1} investigated general MA architectures and practical implementations, with a focus on array-level MAs.
It also provided an overview of candidate implementation methods for the proposed MA architectures, utilizing either direct mechanical or equivalent electronic control. \cite{6DMA} considered six-dimensional MA where array rotation is considered for communication performance enhancement. Moreover, \cite{MAR2} investigated flexible antenna arrays' potential by evaluating the impact of the flexible DoF including rotatable arrays, bendable arrays, and foldable arrays on the multi-sector sum-rate.

Weighted minimum mean square error (WMMSE), which addresses a matrix-weighted sum-MSE minimization problem by introducing a weight matrix to make it equivalent to sum-rate maximization, delivers exceptional performance in joint precoder and combiner optimization for multi-user multiple-input multiple-output (MU-MIMO) systems \cite{WMMSE1}. Beyond the general scenario, it has been extensively applied in various wireless applications, including self-interference suppression in in-band full-duplex communications \cite{WMMSEFD}, the communication-sensing rate trade-off in ISAC \cite{WMMSEISAC}, and hybrid beamforming in MU-MIMO mmWave systems \cite{HWMMSE}. Furthermore, to unlock additional potential of WMMSE, recent works have aimed to accelerate its computation. For instance, \cite{DWMMSE1,DWMMSE2} leveraged deep learning methods to reduce WMMSE's complexity while maintaining comparable performance, and \cite{RWMMSE} proposed a reduced-WMMSE approach, exploiting the structure of linear precoding schemes to lower complexity when the number of antennas at the BS exceeds the number of data streams.

 All existing works on WMMSE primarily focus on the DoF related to the precoder/combiner, i.e., the antenna coefficients, while neglecting the DoF associated with antenna positions. This raises the question:  \textbf{\emph{what will happened when WMMSE beamforming meets MAs?}} By leveraging the modification of the multi-path effect through MAs, WMMSE beamforming can achieve a higher sum-rate with the joint optimization of both antenna coefficients and positions. This approach is termed flexible WMMSE (F-WMMSE) beamforming, inspired by the adaptability of antenna positions. Our key contributions and innovations are summarized as follows:
 \footnote{\label{note1}The source code of this work is open in \url{https://github.com/YyangSJ/F-WMMSE-beamforming} for readers studying.}:
\begin{itemize}
	\item First, we review the classical WMMSE algorithm by introducing its optimization objective and presenting its alternating solutions for the precoder, combiner, and auxiliary weight matrix. Through further derivations, we unify the objectives of the WMMSE precoder and combiner, transforming them into a regularized least squares (RLS) problem.
	\item Subsequently,  we formulate a compressive sensing or sparse recovery problem for the joint optimization of the precoder/combiner and the transmit/receive antenna positions, based on the unified RLS problem for WMMSE. Specifically, we introduce a virtual channel representation (VCR) by defining a dictionary of candidate antenna positions within the allowed movable region. Each column of this dictionary represents a candidate antenna position sampled by \(L\) channel paths. Thus, our objective is to identify the \(N\) dictionary columns (antenna positions) and their corresponding coefficients (precoder/combiner) under the RLS framework. This results in a regularized sparse optimization problem involving the \(\ell_0\)-norm, and we propose the RLS-based simultaneous orthogonal matching pursuit (RLS-SOMP) algorithm to solve it.
	
	\item Finally, to reduce the complexity of the entire F-WMMSE procedure, we propose the fast RLS-SOMP algorithm. This method reduces matching complexity through pre-calculations, at the cost of increased memory usage, and employs the matrix inverse lemma (MIL) to establish a recursive equation, accelerating the RLS computation by leveraging the result from the previous iteration.
	
\end{itemize}

The rest of this paper is organized as follows: Section \ref{sys} introduces the signal and channel models. Section \ref{RWMMSE} reviews the WMMSE algorithm.
Section \ref{PFWMMSE} reformulates WMMSE problems and introduces the proposed F-WMMSE algorithm. 
Section \ref{TCA} analyzes time complexity of WMMSE and F-WMMSE, and proposes low-complexity F-WMMSE implementation.
Section \ref{SR} provides simulation results to evaluate the proposed methods.
Section \ref{Con} concludes this paper.

{\emph {Notations}}:
  ${\left(  \cdot  \right)}^{ T}$, ${\left(  \cdot  \right)}^{ H}$, and $\left(\cdot\right)^{-1}$ denote   transpose, conjugate transpose, and inverse, respectively. $\vert\cdot\vert$ denotes the cardinality of a set. ${\rm Tr}(\mathbf{B})$ denotes the trace of matrix $\mathbf{B}$.
$\Vert\mathbf{B}\Vert_F$ denotes the Frobenius norm of matrix $\mathbf{B}$.   Let \([\mathbf{B}]_{:,i}\) denote the \(i\)-th column of matrix \(\mathbf{B}\), and \([\mathbf{B}]_{:,\bm{\Lambda}}\), where \(\bm{\Lambda}\) is an index set, denote the submatrix of \(\mathbf{B}\) consisting of columns corresponding to the indices in \(\bm{\Lambda}\). Additionally, \(\mathcal{CN}(\mu, \sigma^2\mathbf{I})\) represents the complex Gaussian distribution with mean \(\mu\) and variance \(\sigma^2\).
 $\mathbb{E}[\cdot]$ denotes the expectation operator.

\section{System Model}\label{sys}
   \begin{figure*}
 	\centering 
 	\includegraphics[width=6.35in]{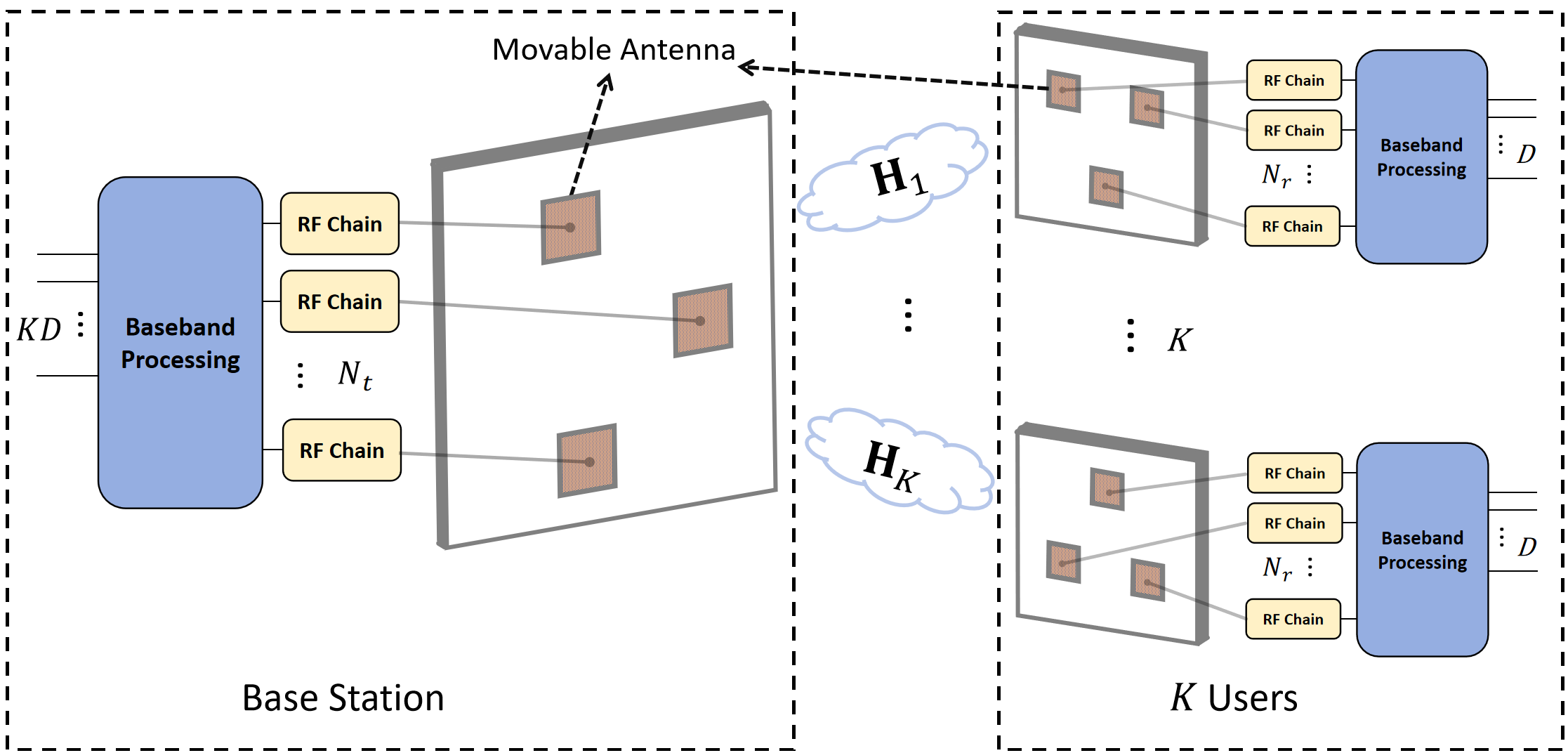}
 	\caption{Description of a MU-MIMO MA system.}\label{syst} 
 \end{figure*} 
  
We consider a MU-MIMO downlink system, where the BS is equipped with \(N_t\) movable antenna elements, distributed along the \(x\)-\(z\) plane, and each of the \(K\) users is equipped with \(N_r\) movable antenna elements. The BS transmits \(D\) data streams to each user, where \(D \leq N_r < N_t\). In this downlink communication system, the received signal at the \(k\)-th user, \(k \in \{1, \cdots, K\}\), is expressed as
\begin{equation}\label{yk}
	\mathbf{y}_k=\mathbf{H}_k\mathbf{F}\mathbf{s}+\mathbf{n}_k,
\end{equation}
where $\mathbf{h}_k\in\mathbb{C}^{N_t\times 1}$ is the $k$-th user's channel, $\mathbf{F}\triangleq[\mathbf{F}_1,\cdots,\mathbf{F}_K]\in\mathbb{C}^{N_t\times DK}$  with $\mathbf{F}_k\in\mathbb{C}^{N_t\times D}$ denoting the precoding matrix for user $k$,
$\mathbf{s}\in\mathbb{C}^{KD\times 1}$ represents the $KD$ data streams for $K$ users, $\mathbf{n}_k$ represents the Gaussian additive white noise following $\mathcal{CN}(0,\sigma^2_k\mathbf{I}_{N_r})$, and $\mathbf{H}_k\in\mathbb{C}^{N_r\times N_t}$ denotes the wireless channel from the BS to user $k$.

 By assuming $L$ paths for all user channels, the spatial geometry channel model is established by, $\forall k$,
 \begin{equation}\label{hk}
 	\begin{aligned}
 		\mathbf{H}_k=&\sqrt{\frac{1}{L}}\sum_{l=1}^{L}\beta_{k,l} \mathbf{a}_r(\theta^r_{k,l},\phi^r_{k,l})\mathbf{a}_t^H(\theta^t_{k,l},\phi^t_{k,l}) \\
 		=& \mathbf{A}_{{\rm R},k}\bm{\Sigma}_k \mathbf{A}_{{\rm T},k}^H,
 	\end{aligned}
 \end{equation}
where $\mathbf{A}_{{\rm R},k}\triangleq [
\mathbf{a}_r(\theta^r_{k,1},\phi^r_{k,1}),
\cdots,\mathbf{a}_r(\theta^r_{k,L},\phi^r_{k,L})
]$,  $\mathbf{A}_{{\rm T},k}\triangleq [
\mathbf{a}_t(\theta^t_{k,1},\phi^t_{k,1}),
\cdots,\mathbf{a}_t(\theta^t_{k,L},\phi^t_{k,L})
]$, and $\bm{\Sigma}_k\triangleq \sqrt{\frac{1}{L}} {\rm diag}\left([\beta_{k,1},\cdots,\beta_{k,L}]\right)
$. Moreover,
$\beta_{k,l}$ is the complex path gain of the $l$-th path of the $k$-th user's channel, $L$ is number of spatial channel paths,
$\phi_{k,l}^{t/r}\triangleq \sin\left(\varphi_{k,l}^{t/r}\right)\sin\left(\vartheta_{k,l}^{t/r}\right)$, and $\theta_{k,l}^{t/r}\triangleq\cos\left(\vartheta_{k,l}^{t/r}\right)$ denote the virtual angle-of-departure (AoD)/angle-of-arrival (AoA) of the $l$-th path of the $k$-th user's channel. Ignoring the subscript, the array-angle manifold $\mathbf{a}(\theta,\phi)\in\mathbb{C}^{N\times 1}$ follows
\begin{equation}\label{aam}
	\begin{aligned}
		\mathbf{a}(\theta,\phi)&=\left[ e^{j\frac{2\pi}{\lambda}(\phi x_1+\theta z_1)}, \cdots, e^{j\frac{2\pi}{\lambda}(\phi x_{N}+\theta z_{N})} \right]^T,
	\end{aligned}
\end{equation}
where $(x_{n},z_{n})$ is the position of the $n$-th antenna in the $x$-$z$ plane.  The set \(\mathbf{p} \triangleq \{(x_{n_t}, z_{n_t}) \mid n_t = 1, \cdots, N_t\}\) denotes the transmit antenna position set, and \(\mathbf{q}_k \triangleq \{(x_{n_{r,k}}, z_{n_{r,k}}) \mid n_{r,k} = 1, \cdots, N_r\}\) represents the receive antenna position set for the \(k\)-th user.

\section{Review of WMMSE}\label{RWMMSE}

We begin by reviewing the WMMSE algorithm, which tackles a matrix-weighted sum-MSE minimization problem to maximize the sum-rate in a MU-MIMO scenario. First, the signal model in Eq. (\ref{yk}) is reformulated by
\begin{equation}
	\mathbf{y}_k= \mathbf{H}_k\mathbf{F}_k\mathbf{s}_k+\sum_{i=1,i\neq k}^K\mathbf{H}_k\mathbf{F}_i\mathbf{s}_i+\mathbf{n}_k,
\end{equation}
where the first term represents the desired signal for user \(k\), while the second term accounts for the multi-user interference from other users. By treating the interference as noise and employing a linear receive beamforming strategy, the estimated signal for each user \(k\) is expressed as follows for all \(k\),
\begin{equation}
	\widehat{\mathbf{s}}_k=\mathbf{W}^H_k\mathbf{y}_k,
\end{equation}
where $\mathbf{W}_k\in\mathbb{C}^{N_r\times D}$ denotes the combiner matrix at user $k$. 

The problem of interest is to jointly optimize the precoder \(\mathbf{F}\) and combiner \(\mathbf{W}_k\), \(\forall k\), to maximize a specific system utility, while ensuring that the total power budget \(P\) is respected. The power constraint is given by
\begin{equation}
	{\rm Tr}\left(\mathbf{F}\mathbf{F}^H
	\right)\leq P.
\end{equation}

A commonly studied utility maximization problem is the weighted sum-rate (WSR) maximization, which can be formulated as:
\begin{equation}\label{wsr}
\begin{aligned}
	&\underset{\mathbf{F}}{{\rm arg \ max}} \ \sum_{k=1}^{K} \alpha_k R_k, \\
	& {\rm s.t.} \	{\rm Tr}\left(\mathbf{F}\mathbf{F}^H
	\right)\leq P,
\end{aligned}
\end{equation}
where $\alpha_k$ represents  the priority of user $k$, and $R_k$ is the rate of user $k$, given by
\begin{equation}\label{Rk}
	\begin{aligned}
	 R_k= 	\log\det\left(\mathbf{I}_{N_r}+\mathbf{C}^{-1}\mathbf{H}_k\mathbf{F}_k\mathbf{F}_k^H\mathbf{H}_k^H
		\right),
	\end{aligned}
\end{equation}
where $\mathbf{C}\triangleq\sum_{i\neq k}\mathbf{H}_i\mathbf{F}_i\mathbf{F}_i^H\mathbf{H}_i^H+\sigma_k^2\mathbf{I}_{N_r}$.

In addition, another popular utility maximization problem is sum-MSE minimization. The MSE matrix $\mathbf{E}_k$ of user $k$ is expressed as
\begin{equation}
	\begin{aligned}
	\mathbf{E}_k =&\mathbb{E}\left[(\mathbf{s}_k-\widehat{\mathbf{s}}_k)(\mathbf{s}_k-\widehat{\mathbf{s}}_k)^H
		\right] \\
		=& \mathbf{W}_k^H\mathbb{E}\left[\mathbf{y}_k\mathbf{y}_k^H\right]\mathbf{W}_k-\mathbb{E}\left[\mathbf{s}_k\mathbf{y}_k^H\right]\mathbf{W}_k-\mathbf{W}_k^H\mathbb{E}\left[\mathbf{y}_k\mathbf{s}_k^H\right]\\
		& +\mathbb{E}\left[\mathbf{s}_k\mathbf{s}_k^H\right]
		 \\ 
		= & \mathbf{W}^H_k \mathbf{H}_k\mathbf{F}\mathbf{F}^H\mathbf{H}_k^H \mathbf{W}_k+ \sigma_k^2\mathbf{W}^H_k\mathbf{W}_k- \mathbf{F}_k^H\mathbf{H}_k^H\mathbf{W}_k\\
		& -\mathbf{W}_k^H\mathbf{H}_k\mathbf{F}_k+\mathbf{I}_{D}
		 \\
		=& \left(\mathbf{I}_{D}- \mathbf{W}_k^H\mathbf{H}_k\mathbf{F}_k
		\right)\left(\mathbf{I}_{D}- \mathbf{W}_k^H\mathbf{H}_k\mathbf{F}_k
		\right)^H \\
		&+\sum_{j\neq k}\mathbf{W}_k^H\mathbf{H}_k\mathbf{F}_j\mathbf{F}_j^H\mathbf{H}_k^H\mathbf{W}_k+ \sigma_k^2\mathbf{W}^H_k\mathbf{W}_k.
	\end{aligned}
\end{equation}
Then, the sum-MSE minimization problem is formulated by
\begin{equation}
\begin{aligned}
	&\underset{\mathbf{F},\mathbf{W}_k}{{\rm arg \ min}} \ \sum_{k=1}^{K} {\rm Tr}\left(\mathbf{E}_k\right) \\ 	&\  {\rm s.t.}\ {\rm Tr}\left(\mathbf{F}\mathbf{F}^H
	\right)\leq P.
\end{aligned}
\end{equation}

Given the above two utility problem,
matrix-weighted sum-MSE minimization establishes their relationship. Let $\mathbf{B}_k\succeq 0 \in\mathbb{C}^{D\times D}$ be the auxiliary weight matrix, \cite{WMMSE1} has shown that problem (\ref{wsr}) is equivalent to 
\begin{equation}\label{MSMSE1}
\begin{aligned}
\underset{\mathbf{F},\mathbf{B}_k,\mathbf{W}_k}{\rm arg \ min} & \sum_{k=1}^{K} \alpha_k \left( {\rm Tr}\left(\mathbf{B}_k\mathbf{E}_k\right)-\log \det \left(\mathbf{B}_k\right)
\right) \\  &  {\rm s.t.}\ {\rm Tr}\left(\mathbf{F}\mathbf{F}^H
\right)\leq P,
\end{aligned}
\end{equation}
in the sense that the global optimal solution \(\mathbf{F}\) is the same for both problems.

Therefore, sum-rate maximization in MU-MIMO can be achieved by solving problem (\ref{MSMSE1}), which employs the block coordinate descent method to iteratively derive the closed-form expressions for \(\{\mathbf{B}, \mathbf{W}, \mathbf{F}\}\).
 \cite{DWMMSE1} demonstrated that the power constraint can be integrated into the objective of problem (\ref{wsr}). This approach allows for solving the problem without the maximum transmit power constraint, followed by scaling the results to comply with the power constraint. In this context, the equivalent rate expression $\widetilde{R}_k$ for user $k$ incorporates the power constraint as follows:
\begin{equation}\label{wR}
	\begin{aligned}
			 \widetilde{R}_k=  	\log\det&\left(\mathbf{I}_{N_r}+\mathbf{H}_k\mathbf{F}_k\mathbf{F}_k^H\mathbf{H}_k^H\left(\sum_{i\neq k}\mathbf{H}_i\mathbf{F}_i\mathbf{F}_i^H\mathbf{H}_i^H \right. \right. \\ & \ \ \left. \left.
			 +\frac{\sigma^2_k}{P} {\rm Tr}\left(
			 \mathbf{F}\mathbf{F}^H\right)\mathbf{I}_{N_r}
	\right)^{-1}
	\right).
	\end{aligned}
\end{equation}

Furthermore, the equivalent MSE matrix for user $k$ is given by
\begin{equation}\label{wEk}
	\begin{aligned}
		\widetilde{\mathbf{E}}_k 
		=& \left(\mathbf{I}_D- \mathbf{W}_k^H\mathbf{H}_k\mathbf{F}_k
		\right)\left(\mathbf{I}_D- \mathbf{W}_k^H\mathbf{H}_k\mathbf{F}_k
		\right)^H \\
		+&\sum_{j\neq k}\mathbf{W}_k^H\mathbf{H}_k\mathbf{F}_j\mathbf{F}_j^H\mathbf{H}_k^H\mathbf{W}_k+   \frac{\sigma^2_k}{P} {\rm Tr}\left(
		\mathbf{F}\mathbf{F}^H\right)
		\mathbf{W}^H_k\mathbf{W}_k.
	\end{aligned}
\end{equation}

Then, we can solve the following equivalent WMMSE problem for  precoder and combiner optimization:
\begin{equation}\label{MSMSE2}
		\underset{\mathbf{F},\mathbf{B},\mathbf{W}}{\rm arg \ min}   \sum_{k=1}^{K} \alpha_k \left( {\rm Tr}\left(\mathbf{B}_k\widetilde{\mathbf{E}}_k\right)-\log \det \left(\mathbf{B}_k\right)
		\right).
\end{equation}

Compared to solving problem (\ref{MSMSE1}), addressing the above problem does not require consideration of the power constraint for objective minimization. It has been shown that their solutions are equivalent \cite{DWMMSE1}. Furthermore, the alternative solutions are provided below and summarized in Algorithm \ref{WMMSE}.

\begin{algorithm} 
	\caption{The procedure of WMMSE}\label{WMMSE} 
	\KwData {Iterative number $\mathcal{I}$, total power $P$, user priority $\{\alpha_k\}_{k=1}^K$, and noise power $\{\sigma_k^2\}_{k=1}^K$.
	}
	\KwResult {Precoder $\mathbf{F}^\star$ and combiner $\{\mathbf{W}_k^\star\}_{k=1}^K$.}
	\BlankLine
	\Begin{ 
		$\textbf{Initialization:}$ Generate random $\mathbf{F}^\star$ satisfying ${\rm Tr}\left(\mathbf{F}^\star\mathbf{F}^{\star,H}
		\right)\leq P$.
		\\  
		\For{$i=1,\cdots,\mathcal{I}$}{ 	\textbf{----------------- \emph{UPDATE $\mathbf{W}_k,\forall k$} ---------------}
			\\
			$\gamma_{1,k}=\frac{\sigma^2_k}{P} {\rm Tr}\left(
			\mathbf{F}^{\star}\mathbf{F}^{\star,H}\right)$
			\; 
			$\mathbf{W}_k^\star=\left( \mathbf{H}_k\mathbf{F}^{\star}\mathbf{F}^{\star,H}\mathbf{H}_k^H+\gamma_{1,k}  \mathbf{I}_{N_r}
			\right)^{-1} \mathbf{H}_k\mathbf{F}^{\star}_k$\;
			\textbf{--------------- \emph{UPDATE $\mathbf{B}_k,\forall k$}--------------------}
			\\
			${\mathbf{B}}^\star_k=\left(\mathbf{I}_D- \mathbf{F}_k^{\star,H}\mathbf{H}_k^H\mathbf{W}^{\star}_k\right)^{-1}$\;
			\textbf{----------------- \emph{UPDATE $\mathbf{F}_k,\forall k$}------------------}\\ $\gamma_2=\sum_{k}\frac{\alpha_k \sigma_k^2}{P}{\rm Tr}\left( 	 \mathbf{W}_k^\star \mathbf{B}_k^\star\mathbf{W}^{\star,H}_k\right)$\; 
		$\mathbf{F}^\star_k=\alpha_k  (\sum_{k}
		\alpha_k \mathbf{H}_k^H \mathbf{W}_k^\star \mathbf{B}_k^\star \mathbf{W}^{\star,H}\mathbf{H}_k\newline {} \ \ \ \ \ \  \ \ \ \ \ \  +  \gamma_2 \mathbf{I}_{N_t}
	)^{-1}\mathbf{H}^H_k\mathbf{W}_k^\star\mathbf{B}_k^\star$\;
		}  
		Normalize $\mathbf{F}^\star\leftarrow \sqrt{\frac{P}{{\rm Tr}\left(\mathbf{F}^\star\mathbf{F}^{\star,H}\right)}}\mathbf{F}^\star$
	} 
\end{algorithm}
\subsection{Update $\mathbf{W}_k,\forall k$ in WMMSE}

 The combiner optimization in the WMMSE problem can be interpreted as a MMSE receiver problem, which can be solved by
\begin{equation}\label{MMSE1}
\mathbf{W}_k^\star=\left( \mathbf{H}_k\mathbf{F}\mathbf{F}^H\mathbf{H}_k^H+\frac{\sigma^2_k}{P} {\rm Tr}\left(
\mathbf{F}\mathbf{F}^H\right)\mathbf{I}_{N_r}
\right)^{-1} \mathbf{H}_k\mathbf{F}_k.
\end{equation}

This can be derived by two different approaches: 1) using the orthogonality principle such that $ \mathbb{E}\left[ (\mathbf{s}_k-\widehat{\mathbf{s}}_k)\mathbf{y}^H
\right]=\mathbf{0}$, and 2) setting the partial derivative of $\mathbf{W}_k$ to zero.
\subsection{Update $\mathbf{B}_k,\forall k$ in WMMSE}
By substituting \(\mathbf{W}^\star_k\) into the equivalent MSE matrix \(\widetilde{\mathbf{E}}_k\) in Eq. (\ref{wEk}), the expression simplifies to
\begin{equation}
	\begin{aligned}
	&	\widetilde{\mathbf{E}}_k ^\star\\ =& 
		 \mathbf{I}_D -\mathbf{W}_k^{\star,H}\mathbf{H}_k\mathbf{F}_k-\mathbf{F}_k^H\mathbf{H}_k^H\mathbf{W}^\star_k +\mathbf{W}_k^{\star,H}\mathbf{H}_k\mathbf{F}_k \mathbf{F}_k^H\mathbf{H}_k^H\mathbf{W}_k^\star \\ 
			&\ 	+\sum_{j\neq k}\mathbf{W}_k^{\star,H}\mathbf{H}_k\mathbf{F}_j\mathbf{F}_j^H\mathbf{H}_k^H\mathbf{W}_k^\star+   \frac{\sigma^2_k}{P} {\rm Tr}\left(
		\mathbf{F}\mathbf{F}^H\right)
		\mathbf{W}^{\star,H}_k\mathbf{W}_k^\star \\
	=	& \mathbf{I}_D -\mathbf{W}_k^{\star,H}\mathbf{H}_k\mathbf{F}_k-\mathbf{F}_k^H\mathbf{H}_k^H\mathbf{W}_k ^\star \\ 
		&\ \ 	+ \mathbf{W}_k^{\star,H}\left(\mathbf{H}_k\mathbf{F}\mathbf{F}^H\mathbf{H}_k^H+\frac{\sigma^2_k}{P} {\rm Tr}\left(
		\mathbf{F}\mathbf{F}^H\right)\mathbf{I}_{N_r} \right)\mathbf{W}_k^\star \\
		\overset{(a)}{=}& \mathbf{I}_D-\mathbf{F}_k^H\mathbf{H}_k^H\mathbf{W}_k^{\star}, 
	\end{aligned}
\end{equation}
 where $(a)$ follows from substituting $\mathbf{W}_k^\star$ in Eq. (\ref{MMSE1}) into the calculations. Notably, some works prefer to express $	\widetilde{\mathbf{E}}_k ^\star=\mathbf{I}_D-\mathbf{W}_k^{\star,H}\mathbf{H}_k\mathbf{F}_k$, which is also acceptable since $\mathbf{F}_k^H\mathbf{H}_k^H\mathbf{W}_k^{\star}$ is a Hermitian matrix.

Considering problem (\ref{MSMSE1}), with the precoder and combiner variables fixed, the subproblem with respect to \(\mathbf{B}_k\) is convex, and the optimal solution is given by
\begin{equation}
	\begin{aligned}
		\mathbf{B}_k^\star & = \left( \widetilde{\mathbf{E}}_k^\star\right)^{-1}.  
	\end{aligned}
\end{equation}
\subsection{Update $\mathbf{F}_k,\forall k$ in WMMSE}
Similarly, the subproblem for the precoder \(\mathbf{F}_k\), with the other variables fixed, can be solved as follows:
\begin{equation}\label{MF2}
	\underset{\mathbf{F}}{\rm arg \ min}    \sum_{k=1}^{K} \alpha_k  {\rm Tr}\left(\mathbf{B}_k\widetilde{\mathbf{E}}_k\right).
\end{equation}

 The first-order optimality condition is satisfied, and the solution for \(\mathbf{F}^{\star}_k\) is given by 
\begin{equation}\label{Fss}
	\begin{aligned}
		\mathbf{F}^{\star}_k=\alpha_k &\left(\sum_{k}
		\alpha_k \mathbf{H}_k^H \mathbf{W}_k \mathbf{B}_k \mathbf{W}^H_k\mathbf{H}_k \right. \\ & \left. +\sum_{k}\frac{\alpha_k\sigma_k^2}{P}{\rm Tr}\left(\mathbf{W}_k\mathbf{B}_k\mathbf{W}_k^H
		\right) \mathbf{I}_{N_t}
		\right)^{-1}\mathbf{H}^H_k\mathbf{W}_k\mathbf{B}_k,
	\end{aligned}
\end{equation}
followed by scaling with the factor $\sqrt{\frac{{P}}{{\rm Tr}\left( \left( \mathbf{F}^{\star }\right)^H\mathbf{F}^{\star  } 
		\right)}}$.

\section{Proposed F-WMMSE}\label{PFWMMSE}
By considering the additional DoF provided by the channel in MA systems, the WSR maximization problem becomes
\begin{equation}\label{wsr2}
	\begin{aligned}
		&\underset{\mathbf{F}, \mathbf{p},\mathbf{q}_{k}}{{\rm arg \ max}} \ \sum_{k=1}^{K} \alpha_k \widetilde{R}_k,  \\
		&{\rm s.t.}  \left\vert [\mathbf{p}]_i-[\mathbf{p}]_j
		 \right\vert\geq d_{\rm min}, \mathbf{p}\in\mathcal{U}_t, \\
		 &\ \ \ \  \left\vert [\mathbf{q}_k]_i-[\mathbf{q}_k]_j
		 \right\vert\geq  d_{\rm min}, \mathbf{q}_k\in\mathcal{U}_r, \forall k,
	\end{aligned}
\end{equation}
where $\mathbf{p}$ represents the antenna position set of the BS, 
and $\mathbf{q}_{k}$ represents the antenna position set of user $k$. The second and third constraints 
enforce that the inter-element spacing between antennas must be greater than a minimum allowable distance \(d_{\rm min}\), and the antenna positions of both the BS and the users are restricted to feasible regions \(\mathcal{U}_t\) and \(\mathcal{U}_r\), respectively.
These regions are defined as 
\begin{equation}\label{Ut}
\mathcal{U}_t=\left\{ (x_{n_t},z_{n_t}) | x_{n_t}\in [0,U_{t}], z_{n_t}\in[0,U_{t}]
\right\},
\end{equation}
\begin{equation}\label{Ur}
	\mathcal{U}_r=\left\{ (x_{n_r},z_{n_r}) | x_{n_r}\in [0,U_{r}], z_{n_r}\in[0,U_{r}]
	\right\},
\end{equation}
where we assume square regions with side lengths \(U_t\) and \(U_r\) for the BS and users, respectively.

Compared to problem (\ref{wsr}), the WSR problem in MA systems further incorporates the optimization of antenna positions \(\{\mathbf{p}, \mathbf{q}_k, \forall k\}\), which influence the channel \(\mathbf{H}_k\) and consequently affect the achievable rate in Eq. (\ref{wR}). In this sense, the globally optimal solution for the WSR problem remains equivalent to the matrix-weighted sum-MSE minimization problem given by
\begin{equation}\label{FM}
	\begin{aligned}
		\underset{\mathbf{F},\mathbf{p},\mathbf{B}_k,\mathbf{W}_k,\mathbf{q}_k}{\rm arg \ min} & \sum_{k=1}^{K} \alpha_k \left( {\rm Tr}\left(\mathbf{B}_k\widetilde{\mathbf{E}}_k\right)-\log \det \left(\mathbf{B}_k\right)
		\right) \\  &  {\rm s.t.}  \left\vert [\mathbf{p}]_i-[\mathbf{p}]_j
		\right\vert\geq d_{\rm min}, \mathbf{p}\in\mathcal{U}_t, \\
		&\ \ \ \  \left\vert [\mathbf{q}_k]_i-[\mathbf{q}_k]_j
		\right\vert\geq  d_{\rm min}, \mathbf{q}_k\in\mathcal{U}_r, \forall k.
	\end{aligned}
\end{equation}

In the following, we derive alternative solutions for F-WMMSE in the context of problem (\ref{FM}).
\subsection{Update $\{\mathbf{W}_k,\mathbf{q}_k,\forall k\}$ in F-WMMSE
}
This subsection jointly optimizes $\mathbf{W}_k$ and $\mathbf{q}_k$.
By fixing $\{\mathbf{F},\mathbf{p},\mathbf{B}_k\}$, minimizing the weighted sum-MSE leads to the F-MMSE receiver problem:
\begin{equation}\label{P_MMSE}
 \begin{aligned}
 &\underset{\mathbf{W}_k,\mathbf{q}_k}{\rm arg \ min} \ \mathbb{E}\left[ \left\Vert \mathbf{s}_k-\widehat{\mathbf{s}}_k \right\Vert_2^2
 \right] \\
 &{\rm s.t.} \ \left\vert [\mathbf{q}_k]_i-[\mathbf{q}_k]_j
 \right\vert\geq  d_{\rm min}, \mathbf{q}_k\in\mathcal{U}_r, \forall k.
 \end{aligned}
\end{equation}

It is challenging to jointly solve for \(\mathbf{W}_k\) and \(\mathbf{q}_k\) from the above problem. The following proposition reformulates the objective.

\begin{Proposition}\label{pro_1}
The multi-user MMSE receiver problem is equivalent to the following Frobenius norm regularized least squares problem as
\begin{equation}\label{FW1}
\underset{\mathbf{W}_k}{\rm arg \ min} \
\left\Vert \widetilde{\mathbf{I}}_k -\mathbf{F}^H\mathbf{H}_k^H \mathbf{W}_k
\right\Vert_F^2 + \frac{\sigma^2_k}{P} {\rm Tr}\left(
\mathbf{F}\mathbf{F}^H\right)\left\Vert\mathbf{W}_k\right\Vert_F^2,
\end{equation}
where $\widetilde{\mathbf{I}}_k\triangleq
\begin{bmatrix}
\mathbf{0}_{D\times (k-1)D}, \mathbf{I}_D, \mathbf{0}_{D\times (K-k)D}
\end{bmatrix}^T\in\mathbb{C}^{KD\times K}$ represents the $k$-th user's indicator. Its solution is given by
\begin{equation}\label{FW1_S}
 \mathbf{W}_k=\left(\mathbf{H}_k\mathbf{F}\mathbf{F}^H\mathbf{H}_k^H+\frac{\sigma^2_k}{P} {\rm Tr}\left(
 \mathbf{F}\mathbf{F}^H\right)\mathbf{I}_{N_r}\right)^{-1}\mathbf{H}_k\mathbf{F}\widetilde{\mathbf{I}}_k,
\end{equation}
which is equivalent to Eq. (\ref{MMSE1}) as $\mathbf{F}_k=\mathbf{F}\widetilde{\mathbf{I}}_k$.
\end{Proposition}
\begin{Proof}
Please refer to  Appendix \ref{W_bian}.	
\end{Proof}

\begin{remark}
	\emph{The subproblem of the WMMSE combiner in problem (\ref{FW1}) is equivalent to the regularized zero-forcing equalization problem with the regularized factor $\frac{\sigma_k^2}{P}{\rm Tr}\left(\mathbf{F}\mathbf{F}^H\right)$ for user $k$. 
	}
\end{remark}

Given Proposition \ref{pro_1}, the problem (\ref{P_MMSE}) can be reformulated as 
\begin{equation}\label{CS1}
	\begin{aligned}
	&	\underset{\mathbf{W}_k,\mathbf{q}_k}{\rm arg \ min} \
		\left\Vert \widetilde{\mathbf{I}}_k -\mathbf{F}^H\mathbf{H}_k^H \mathbf{W}_k
		\right\Vert_F^2 + \frac{\sigma^2_k}{P} {\rm Tr}\left(
		\mathbf{F}\mathbf{F}^H\right)\left\Vert\mathbf{W}_k\right\Vert_F^2 \\
 &\ \ \ \ \ \ {\rm s.t.} \ \left\vert [\mathbf{q}_k]_i-[\mathbf{q}_k]_j
\right\vert\geq  d_{\rm min}, \mathbf{q}_k\in\mathcal{U}_r, \forall k.
\end{aligned} 
\end{equation}

This problem remains difficult to solve due to the non-linear dependence on the antenna positions. Notably, antenna position optimization can be viewed as a sparse optimization problem, where the antenna positions are initially treated as discrete variables, enabling on-grid atom selection for joint optimization of the antenna coefficients and positions. 

First, recalling $\mathbf{H}_k^H=\mathbf{A}_{{\rm T},k}\bm{\Sigma}_k^H\mathbf{A}_{{\rm R},k}^H$ in Eq. (\ref{hk}),
 we design a dictionary or position codebook $\mathbf{G}_{{\rm R},k}\in\mathbb{C}^{L\times G_r}$ to sample $\mathbf{A}_{{\rm R},k}^H\in\mathbb{C}^{L\times N_r}$ for atom selection, shown as
\begin{equation}\label{Gr}
	{\mathbf{G}}_{{\rm R},k}= \left[\mathbf{g}_{r,k}(x_1,z_1),\cdots,\mathbf{g}_{r,k}(x_{G_r},z_{G_r})\right],
\end{equation}
where $G_r \gg N_r$ denotes the number of atoms or candidate positions, and
\begin{equation}\label{apm}
	\mathbf{g}_{r,k}(x,z)=\left[e^{-j\frac{2\pi}{\lambda}(\phi_{k,1}^rx+\theta^r_{k,1}z)
	},\cdots, e^{-j\frac{2\pi}{\lambda}(\phi^r_{k,L}x+\theta^r_{k,L}z)}\right]^T
\end{equation}
denotes the array-position manifold. The antenna position set $\{(x_1,z_1),\cdots,(x_{G_r},z_{G_r})\}$
can be regarded as a $G_r$-element virtual array. We assume this virtual array is deployed on a movable region $\mathcal{U}_r$ in Eq. (\ref{Ur}) with half-wavelength spacing such that $\frac{\lambda}{2}\sqrt{G_r}=U_r$,
   allowing us not to consider the inter-element spacing constraint in problem (\ref{CS1}).

 Compared to the array-angle manifold in Eq. (\ref{aam}) which is a vector function with angle as the variable and the number of antennas as its dimension, the array-position manifold in Eq. (\ref{apm}) is a vector function with antenna position as the variable and the number of angle paths as its dimension. 
 Thus, beamspace-domain techniques usually use array-angle manifolds such as angle and channel estimation. Similarly, the array-position manifold introduced in this paper can facilitate antenna position optimization/estimation.  
  Therefore, we establish the VCR with virtual arrays for user $k$,
\begin{equation}
	\mathbf{H}_k^H \overset{\rm VCR}{\Longrightarrow} \mathbf{A}_{{\rm T},k}\bm{\Sigma}_k^H \mathbf{G}_{{\rm R},k}.
\end{equation}

Using the $\ell_0$-norm to constrain the number of antennas, the F-WMMSE combiner problem can be formulated by a regularized sparse recovery problem:
	\begin{equation}\label{WIP}
		\begin{aligned}
			&\underset{\overline{  \mathbf{W}}_k}{\rm arg \ min} \ \left\Vert \widetilde{\mathbf{I}}_k-\bm{\Phi}_{{\rm R},k}\overline{  \mathbf{W}}_k
			\right\Vert_F^2+ \gamma_{1,k}
			\left\Vert\overline{  \mathbf{W}}_k\right\Vert_F^2 \\
			& \ \ \ \ \ \ \ \ \ \ \ {\rm s.t.} \ \left\Vert\overline{  \mathbf{W}}_k\right\Vert_{\rm row,0} =N_r,
		\end{aligned}
	\end{equation} 
where 
$\bm{\Phi}_{{\rm R},k}\triangleq \mathbf{F}^H
\mathbf{A}_{{\rm T},k}\bm{\Sigma}_k^H {\mathbf{G}}_{{\rm R},k}\in\mathbb{C}^{KD\times G_r}$ denotes the sensing matrix, $\gamma_{1,k}\triangleq\frac{\sigma^2_k}{P} {\rm Tr}\left(
\mathbf{F}\mathbf{F}^H\right)$,  $\overline{\mathbf{W}}_k\in\mathbb{C}^{G_r\times D}$ denotes the sparse combiner of user $k$ in which the re-organized non-zero rows  denotes the F-WMMSE combiner and non-zero row indices denote the antenna positions, and 
$\left\Vert \cdot \right\Vert_{{\rm row},0}$ returns the number of non-zero rows of a matrix.

To solve the above sparse optimization problem, the following proposition is developed based on the greedy pursuit method.
\begin{Proposition}\label{RLSSOMP}
	Consider a regularized sparse optimization problem as follows,
	\begin{equation}\label{CSS}
		\begin{aligned}
			&\underset{\mathbf{X}}{\rm arg \ min} \ \left\Vert\mathbf{Y}-\mathbf{D}\mathbf{X}
			\right\Vert_F^2+ \zeta
			\left\Vert\mathbf{X}\right\Vert_F^2 \\
			& {\rm s.t.} \ \left\Vert\mathbf{X}\right\Vert_{\rm row,0} =N.
		\end{aligned}
	\end{equation}
It can be effectively addressed using the RLS-SOMP method outlined in Algorithm \ref{RLS-SOMP}, which replaces the conventional LS computation in the residual updating step with RLS.
\end{Proposition}
\begin{Proof}
Please refer to Appendix \ref{GP}.
\end{Proof}

Based on Proposition \ref{RLSSOMP}, problem (\ref{WIP}) can be directly solved. We can obtain the optimized $N_r$-row-sparse matrix $\overline{\mathbf{W}}_k^\star$, 
 where the $N_r$ non-zero rows denote the $N_r$ antenna's coefficients across $D$ data streams, and
 the $N_r$ non-zero row indices, added to the set $\bm{\Lambda}_{{\rm R},k}$, indicates the selection of $N_r$ columns in dictionary $\mathbf{G}_{{\rm R},k}$, implying $N_r$ antenna positions.

Notably, using Proposition \ref{RLSSOMP} to solve this problem allows for only discrete optimization of antenna positions, which may fall short of achieving truly optimal solutions. This limitation, known as the mismatch problem in recovery theory, arises from the finite size of the dictionary. Several approaches, such as perturbation-based methods \cite{ars2}, the Newton method \cite{newt}, and spatial alternating generalized expectation maximization   \cite{SAGE}, address this issue by incorporating off-grid techniques that refine antenna positions based on the initial on-grid results, further reducing the objective. For simplicity, off-grid refinement is not considered in this work and is left for future exploration.

\begin{algorithm} 
	\caption{RLS-SOMP$(\mathbf{Y}, \mathbf{D}, \zeta)$.}\label{RLS-SOMP}
	\KwData {Measurement signal $\mathbf{Y}$, sensing matrix $\mathbf{D}$, and regularization factor $\zeta$.}
	\KwResult {Coefficient matrix $\mathbf{X}^\star$ and support $\bm{\Lambda}$.}
	\BlankLine
	\Begin{ 
		$\textbf{Initialization:}$ $\mathbf{R}=\mathbf{Y}$, $\bm{\Lambda}=\emptyset$, $\bm{\Gamma}=\{1,\cdots,G\}$.\\
		\For{$n=1,\cdots,N$}{  $g^\star=\underset{g\in\bm{\Gamma}}{{\rm arg \ max}} \ \left\Vert   [\mathbf{D}]_{:,g} ^{H} {\mathbf{R}}  \right\Vert_2^2$ \; 
			$\bm{\Lambda}\leftarrow \bm{\Lambda}\cup g^\star$, $\bm{\Gamma}\leftarrow \bm{\Gamma}\setminus  g^\star$\; 
			$\mathbf{X}^\star=\left([\mathbf{D}]_{:,\bm{\Lambda}}^{H}[\mathbf{D}]_{:,\bm{\Lambda}}+\zeta \mathbf{I}_n\right)^{-1}[\mathbf{D}]_{:,\bm{\Lambda}}^{H}\mathbf{Y}$ \;		$\mathbf{R}=\mathbf{Y}-[\mathbf{D}]_{:,\bm{\Lambda}}\mathbf{X}^\star$\;
		}  
	}	 
\end{algorithm}

\subsection{Update $\mathbf{B}_k,\forall k$ in F-WMMSE}
After determining the optimal receive antenna positions $\mathbf{q}k^\star$ for all $k$ by solving problem (\ref{WIP}), we update the corresponding receive array-angle manifold matrix to $\mathbf{A}_{{\rm R},k}^\star$ using $\mathbf{q}_k^\star$ in accordance with Eq. (\ref{aam}). By substituting $\mathbf{A}_{{\rm R},k}^\star$ and $\mathbf{W}^\star_k$ into the equivalent MSE matrix $\widetilde{\mathbf{E}}_k$ as defined in Eq. (\ref{wEk}), we obtain the solution for the subproblem related to $\mathbf{B}_k$ within problem (\ref{FM}):
\begin{equation} 
	 {\mathbf{B}}^\star_k=\left( \mathbf{I}_D- 
	 \mathbf{F}_k^{H}\mathbf{A}_{{\rm T},k}\bm{\Sigma}_k^H
	 {\mathbf{A}}^{\star,H}_{{\rm R},k}\mathbf{W}_k^{\star}
	 \right)^{-1}.
\end{equation}
\subsection{Update $\{\mathbf{F},\mathbf{p}\}$ in F-WMMSE}
This subsection jointly optimizes $\mathbf{F}$ and $\mathbf{p}$. By fixing $\{\mathbf{W}_k,\mathbf{q}_k,\mathbf{B}_k\}$, the subproblem regarding $\{\mathbf{F},\mathbf{p}\}$ is written as
\begin{equation}\label{FBE}
	\begin{aligned}
		&\underset{\mathbf{F},\mathbf{p}}{\rm arg \ min} \ \sum_{k=1}^K\alpha_k {\rm Tr}\left(\mathbf{B}_k\widetilde{\mathbf{E}}_k\right) \\
		& {\rm s.t.} \	  \left\vert [\mathbf{p}]_i-[\mathbf{p}]_j
		\right\vert\geq d_{\rm min}, \mathbf{p}\in\mathcal{U}_t.
	\end{aligned}
\end{equation}

To address the aforementioned issue, we reformulate the objective in problem (\ref{FBE}). This transformation converts the subproblem of the WMMSE precoder into a regularized least squares problem, as demonstrated in the following proposition.
\begin{Proposition}\label{F_pro}
	Let  $\widetilde{\mathbf{H}}\triangleq \left[ \widetilde{\mathbf{H}}_1^T,\cdots, \widetilde{\mathbf{H}}_K^T
	\right]^T\in\mathbb{C}^{KD\times N_t}
	$ with $\widetilde{\mathbf{H}}_k\triangleq \sqrt{\alpha_k}\mathbf{B}^{\frac{1}{2}}_k \mathbf{W}_k^H\mathbf{H}_k\in\mathbb{C}^{D\times N_t}$, $\widetilde{\mathbf{B}}\triangleq
	{\rm blkdiag}\left( \alpha_1\mathbf{B}_1,\cdots, \alpha_K\mathbf{B}_K
	\right)$,
	and $\gamma_2\triangleq\sum_{k}\frac{\alpha_k \sigma_k^2}{P}{\rm Tr}\left( 	 \mathbf{W}_k \mathbf{B}_k\mathbf{W}^H_k\right)$,
the subproblem of WMMSE precoder is equivalent to a Frobenius norm regularized least squares problem
\begin{equation}\label{FW-F1}
\underset{\mathbf{F}}{\rm arg \ min} \ \left\Vert \widetilde{\mathbf{B}}^{\frac{1}{2}}-\widetilde{\mathbf{H}}\mathbf{F}
\right\Vert_F^2+ \gamma_2
\left\Vert \mathbf{F}\right\Vert_F^2.
\end{equation}
  Its solution is given by
\begin{equation}\label{FW-F2}
	\mathbf{F}=\left( \widetilde{\mathbf{H}}^H \widetilde{\mathbf{H}}+\gamma_2\mathbf{I}_{N_t}\right)^{-1} \widetilde{\mathbf{H}}^H \widetilde{\mathbf{B}}^{\frac{1}{2}},
\end{equation}
which is equivalent to Eq. (\ref{Fss}). Similarly, the scaling factor $\sqrt{\frac{{P}}{{\rm Tr}\left( \mathbf{F}^H\mathbf{F}
		\right)}}$ should be multiplied.
\end{Proposition}

\begin{Proof}
	Please find the proof in Appendix \ref{F_bian}.
\end{Proof}
 \begin{remark}
 {\emph{Proposition \ref{F_pro} establishes the equivalence between problems (\ref{MF2}) and (\ref{FW-F1}),
 		 identifying that the WMMSE precoder can be solved similarly to the combiner, essentially with a same regularized least squares problem. Therefore, we can also use the sparse recovery strategy to address the F-WMMSE precoder problem for joint antenna position and coefficient optimization.}} 
 \end{remark}

 First, according to ${\mathbf{H}}_k=\mathbf{A}_{{\rm R},k}\bm{\Sigma}_k\mathbf{A}_{{\rm T},k}^H$
 we design a dictionary or position codebook $\mathbf{G}_{{\rm T},k}\in\mathbb{C}^{L\times G_t}$ to sample $\mathbf{A}_{{\rm T},k}^H\in\mathbb{C}^{L\times N_r}$ for sparse recovery, shown as
 \begin{equation}\label{Gt}
 	{\mathbf{G}}_{{\rm T},k}= \left[\mathbf{g}_{t,k}(x_1,z_1),\cdots,\mathbf{g}_{t,k}(x_{G_t},z_{G_t})\right],
 \end{equation}
 where $G_t \gg N_t$ denotes the number of atoms or candidate positions, and
 \begin{equation}\label{apm2}
 	\mathbf{g}_{t,k}(x,z)=\left[e^{-j\frac{2\pi}{\lambda}(\phi^t_{k,1}x+\theta^t_{k,1}z)
 	},\cdots, e^{-j\frac{2\pi}{\lambda}(\phi^t_{k,L}x+\theta^t_{k,L}z)}\right]^T.
 \end{equation} 
 Compared to $\mathbf{g}_{r,k}(x,z)$ in Eq. (\ref{apm}) which depends on AoAs at the user end, $\mathbf{g}_{t,k}(x,z)$ depends on AoDs at the BS end.  We assume the $G_t$ elements are deployed on a movable region $\mathcal{U}_t$ in Eq. (\ref{Ut}) with half-wavelength spacing such that $\frac{\lambda}{2}\sqrt{G_t}=U_t$,
 allowing us not to consider the inter-element spacing constraint.
 
 Then, $\widetilde{\mathbf{H}}$ can be re-expressed by VCR: 
 \begin{equation}\label{VCRT}
 	\begin{aligned}
 		\widetilde{\mathbf{H}}
 		=&\begin{bmatrix}
 			\sqrt{\alpha_1}\mathbf{B}^{\frac{1}{2}}_1 \mathbf{W}_1^H\mathbf{A}_{{\rm R},1}\bm{\Sigma}_1\mathbf{A}_{{\rm T},1}^H \\
 			\vdots \\
 			\sqrt{\alpha_K}\mathbf{B}^{\frac{1}{2}}_K \mathbf{W}_K^H\mathbf{A}_{{\rm R},K}\bm{\Sigma}_K\mathbf{A}_{{\rm T},K}^H
 		\end{bmatrix} 	 \\ 
 		\overset{\rm VCR}{\Longrightarrow}& \begin{bmatrix}
 			\sqrt{\alpha_1}\mathbf{B}^{\frac{1}{2}}_1 \mathbf{W}_1^H\mathbf{A}_{{\rm R},1}\bm{\Sigma}_1\mathbf{G}_{{\rm T},1}\\
 			\vdots \\
 			\sqrt{\alpha_K}\mathbf{B}^{\frac{1}{2}}_K \mathbf{W}_K^H\mathbf{A}_{{\rm R},K}\bm{\Sigma}_K\mathbf{G}_{{\rm T},K}
 		\end{bmatrix} \\
 		=& \bm{\Psi} \mathbf{G}_{\rm T},
 	\end{aligned} 
 \end{equation}
 where $\bm{\Psi}\triangleq {\rm blkdiag}\left(   \begin{bmatrix}
 	\sqrt{\alpha_1}\mathbf{B}^{\frac{1}{2}}_1 \mathbf{W}_1^H\mathbf{A}_{{\rm R},1}\bm{\Sigma}_1\\
 	\vdots \\
 	\sqrt{\alpha_K}\mathbf{B}^{\frac{1}{2}}_K \mathbf{W}_K^H\mathbf{A}_{{\rm R},K}\bm{\Sigma}_K
 \end{bmatrix}\right)\in\mathbb{C}^{KD\times KL}$ represents the measurement matrix, and $\mathbf{G}_{{\rm T}}\triangleq \left[\mathbf{G}_{{\rm T},1}^T,\cdots, \mathbf{G}_{{\rm T},K}^T\right]^T\in\mathbb{C}^{KL\times G_t}$ represents the dictionary. 
Denoted by $\bm{\Phi}_{\rm T}\triangleq\bm{\Psi}\mathbf{G}_{\rm T}\in\mathbb{C}^{KD\times G_t}$, the following sparse recovery problem can be formulated for F-WMMSE precoder:
\begin{equation}
	\begin{aligned}
		&\underset{\overline{  \mathbf{F}}}{\rm arg \ min} \ \left\Vert \widetilde{\mathbf{B}}^{\frac{1}{2}}-\bm{\Phi}_{\rm T}\overline{  \mathbf{F}}
		\right\Vert_F^2+ \gamma_2
		\left\Vert\overline{  \mathbf{F}}\right\Vert_F^2 \\
		& {\rm s.t.} \ \left\Vert\overline{  \mathbf{F}}\right\Vert_{\rm row,0} =N_t,
	\end{aligned}
\end{equation} 
where $\overline{\mathbf{F}}\in\mathbb{C}^{G_t\times KD}$ denotes the $N_t$-row-sparse precoder.

The above problem can be addressed by the RLS-SOMP algorithm, described in Proposition \ref{RLSSOMP}. We can obtain the optimized $N_t$-row-sparse matrix $\overline{\mathbf{F}}^\star$, 
where the $N_t$ non-zero rows denote the $N_t$ antenna's coefficients across $KD$ data streams, and
the $N_t$ non-zero row indices, added to the set $\bm{\Lambda}_{{\rm T}}$, indicates the selection of $N_t$ columns in dictionary $\mathbf{G}_{{\rm T}}$, implying $N_t$ antenna positions.

\begin{algorithm} 
	\caption{The procedure of F-WMMSE}\label{F-WMMSE} 
	\KwData {Iterative number $\mathcal{I}$, total power $P$, user priority $\{\alpha_k\}_{k=1}^K$, noise power $\{\sigma_k^2\}_{k=1}^K$, and allowed movable regions $\mathcal{U}_t$ and $\mathcal{U}_r$.}
	\KwResult {Precoder $\mathbf{F}$, combiner $\{\mathbf{W}_k\}_{k=1}^K$, BS's antenna position $\mathbf{p}$, and users' antenna position $\{\mathbf{q}_k\}_{k=1}^K$.}
	\BlankLine
	\Begin{ 
		$\textbf{Initialization:}$ Generate random $\mathbf{F}^\star$ satisfying ${\rm Tr}\left(\mathbf{F}^\star\mathbf{F}^{\star,H}
		\right)\leq P$, construct $\mathbf{G}_{{\rm R},k}$ and $\mathbf{G}_{\rm T}$ according to Eqs. (\ref{Gr}) and (\ref{Gt}).  $\mathbf{A}_{{\rm T},k}^\star=\mathbf{A}_{{\rm T},k},\forall k$.  
		\\  
		\For{$i=1,\cdots,\mathcal{I}$}{ 	\textbf{-------------- \emph{UPDATE $\left\{\mathbf{W}_k,\mathbf{q}_k\right\},\forall k$} ------------}
			\\
			$\gamma_{1,k}=\frac{\sigma^2_k}{P} {\rm Tr}\left(
			\mathbf{F}^{\star}\mathbf{F}^{\star,H}\right)$
			\;
			$	\bm{\Phi}_{{\rm R},k}^\star\triangleq  \mathbf{F}^{\star,H} \mathbf{A}_{{\rm T},k}^\star\bm{\Sigma}_k^H {\mathbf{G}}_{{\rm R},k}$\;
			$[\mathbf{W}_k^\star,\bm{\Lambda}_{{\rm R},k}]=\text{RLS-SOMP}\left(\widetilde{\mathbf{I}}_k,\bm{\Phi}^\star_{{\rm R},k},\gamma_{1,k}\right)$\;	${\mathbf{A}}^\star_{{\rm R},k}= \left[ \mathbf{G}_{{\rm R},k}\right]^H_{:,\bm{\Lambda}_{{\rm R},k}}$\;
			\textbf{--------------- \emph{UPDATE $\mathbf{B}_k,\forall k$}---------------------}
			\\
			${\mathbf{B}}^\star_k=\left( \mathbf{I}_D- 
			\mathbf{F}_k^{\star,H}\mathbf{A}_{{\rm T},k}^{\star}\bm{\Sigma}_k^H
			{\mathbf{A}}^{\star,H}_{{\rm R},k}\mathbf{W}_k^{\star}
			\right)^{-1}$\;
			\textbf{------------------ \emph{UPDATE $\left\{\mathbf{F},\mathbf{p} \right\}$}------------------}\\ $\gamma_2=\sum_{k}\frac{\alpha_k \sigma_k^2}{P}{\rm Tr}\left( 	 \mathbf{W}_k^\star \mathbf{B}_k^\star\mathbf{W}^{\star,H}_k\right)$\;
			$ \bm{\Phi}_{\rm T}^\star= {\rm blkdiag}\left(   \begin{bmatrix}
				\sqrt{\alpha_1}\mathbf{B}^{\star,\frac{1}{2}}_1 \mathbf{W}_1^H\mathbf{A}_{{\rm R},1}^{\star}\bm{\Sigma}_1\\
				\vdots \\
				\sqrt{\alpha_K}\mathbf{B}^{\star,\frac{1}{2}}_K \mathbf{W}_K^H\mathbf{A}_{{\rm R},K}^{\star}\bm{\Sigma}_K
			\end{bmatrix}\right) \mathbf{G}_{\rm T}$
			\;
			$[\mathbf{F}^\star,\bm{\Lambda}_{{\rm T}}]=\text{RLS-SOMP}\left(\widetilde{\mathbf{B}}^{\star,\frac{1}{2}},\bm{\Phi}_{\rm T}^\star,\gamma_2\right)$\;
			${\mathbf{A}}^\star_{{\rm T}}= \left[ \mathbf{G}_{{\rm T}}\right]^H_{:,\bm{\Lambda}_{{\rm T}}}$\;
		}  
		Normalize $\mathbf{F}^\star\leftarrow \sqrt{\frac{P}{{\rm Tr}\left(\mathbf{F}^\star\mathbf{F}^{\star,H}\right)}}\mathbf{F}^\star$
	} 
\end{algorithm}

\subsection{F-WMMSE Algorithm Summary}
Optimizing antenna positions aims to refine the channel conditions. In the spatial channel model considered, optimizing the transmit and receive antenna positions directly impacts the array manifold matrices $\mathbf{A}_{{\rm T},k}$ and $\mathbf{A}_{{\rm R},k},\forall k$, respectively.  This setup allows us to approach F-WMMSE beamforming from the perspective of optimizing   $\mathbf{A}^H_{{\rm T},k}\mathbf{F}$ for precoder and $\mathbf{W}^H_k\mathbf{A}_{{\rm R},k}$ for combiner. However, direct optimization of $\mathbf{A}^H_{{\rm T},k}\mathbf{F}_k$ and $\mathbf{W}^H_k\mathbf{A}_{{\rm R},k}$ 	
is not feasible, as $\mathbf{A} _{{\rm T},k}$ and $\mathbf{A}_{{\rm R},k}$ must adhere to specific array manifold structures.

Compared to the traditional WMMSE method in Algorithm \ref{WMMSE}, which in each iteration $i$, beamforming matrices $\mathbf{F}_k^\star$ and $\mathbf{W}_k^\star$ are updated and used for updating the auxiliary variable $\mathbf{B}_k^\star$,
 F-WMMSE shown in Algorithm \ref{F-WMMSE} additionally updates antenna positions or channel supports for flexible beamforming matrices $\mathbf{A}_{{\rm T},k}^{\star,H}{\mathbf{F}}_k^\star$ and $\mathbf{A}_{{\rm R},k}^{\star,H}{\mathbf{W}}^\star_k$, and also use them to update the auxiliary variable $\mathbf{B}_k^\star$.

\subsection{Simplifications into MU-MISO Case}
Here, we establish the relation between this work and our previous work \cite{FPR} by considering the MU-MISO case. 
In this context, $\mathbf{W}_k$ and $\mathbf{B}_k$ reduce to scalars, the channel matrix $\mathbf{H}_k$ reduces to the vector $\mathbf{h}_k\in\mathbb{C}^{N_t\times 1}$, and the VCR of the multi-user channel in Eq. (\ref{VCRT}) becomes
 \begin{equation} 
	\begin{aligned}
		\widetilde{\mathbf{H}}
		=&\begin{bmatrix}
			\sqrt{\alpha_1}\mathbf{1}\bm{\Sigma}_1\mathbf{A}_{{\rm T},1}^H \\
			\vdots \\
			\sqrt{\alpha_K}\mathbf{1}\bm{\Sigma}_K\mathbf{A}_{{\rm T},K}^H
		\end{bmatrix} 	 \\ 
		\overset{\rm VCR}{\Longrightarrow}& \begin{bmatrix}
			\sqrt{\alpha_1}\bm{\beta}_1^T\mathbf{G}_{{\rm T},1}\\
			\vdots \\
			\sqrt{\alpha_K}\bm{\beta}_K^T\mathbf{G}_{{\rm T},K}
		\end{bmatrix} \\
		=& \bm{\Psi} \mathbf{G}_{\rm T},
	\end{aligned} 
\end{equation}
where $\bm{\Psi}\triangleq {\rm blkdiag}(\sqrt{\alpha_1}\bm{\beta}_1^T
,\cdots,\sqrt{\alpha_K}\bm{\beta}_K^T)$.

	The problem	of WMMSE precoder is equivalent to a Frobenius norm regularized least squares problem
	\begin{equation}\label{FW-F3}
		\underset{\mathbf{F}}{\rm arg \ min} \ \left\Vert \bm{\alpha}^{\frac{1}{2}}-\widetilde{\mathbf{H}}\mathbf{F}
		\right\Vert_F^2+ \frac{\sum_{k}\alpha_k\sigma_k^2}{P}
		\left\Vert \mathbf{F}\right\Vert_F^2,
	\end{equation}
	where $\bm{\alpha}\triangleq {\rm diag}([\alpha_1,\cdots,\alpha_K])$. Assuming all users are fair, i.e., $\alpha_1=\cdots=\alpha_K=1$, the problem simplifies to the flexible regularized zero-forcing (RZF) precoding problem as discussed in \cite{FPR}. This work demonstrates that antenna movement aids in selecting an effective projection subspace for precoding, and that the iterative antenna selection process in the RZF precoding scheme aligns with the RLS-SOMP approach.

 \section{Time Complexity Analysis and Fast Implementation}\label{TCA}
 
 \subsection{Complexity Analysis}
 The time complexity of the WMMSE algorithm can be summarized as follows: Each iteration's computation of $\mathbf{W}_k$
 is primarily determined by the matrix-matrix multiplication $\mathbf{H}_k\mathbf{F}$, which has a complexity of $\mathcal{O}(N_rN_tKD)$. For a total of $K$ users, this results in an overall complexity of $\mathcal{O}(N_rN_tK^2D)$. The calculation of $\{\mathbf{B}_k\}_{k=1}^K$
 incurs a complexity of $\mathcal{O}(KDN_tN_r)$, attributable to the matrix-matrix multiplication $\mathbf{F}_k^H\mathbf{H}_k^H$. Additionally, updating $\mathbf{F}$ through the calculation given in Eq. (\ref{FW-F2}) requires a complexity of $\mathcal{O}(N_t^3)$. Given that $N_t\geq N_r\geq D$, the total complexity of the WMMSE algorithm is dominated by the precoder updating procedure, which can be represented as $\mathcal{O}(\mathcal{I}N_t^3)$, where $\mathcal{I}$ denotes the number of iterations of the WMMSE algorithm.

For F-WMMSE, the complexity is indeed dominated by the precoder updating procedure, which relies on the RLS-SOMP algorithm as outlined in Algorithm \ref{RLS-SOMP}. In this context, given  $\mathbf{Y}\in\mathbb{C}^{KD\times KD}$ and  $\mathbf{D}\in\mathbb{C}^{KD\times G}$, the atom matching process detailed in line 4 incurs a complexity of   $\mathcal{O}(GK^2D^2)$ for each iteration. Additionally, computing the RLS in line 6 requires a complexity of  $\mathcal{O}(n^2KD+n^3)$. When considering the circular loop, the total complexity can be expressed as $\mathcal{O}\left(NGK^2D^2+ \sum_{n=1}^{N}(n^2KD+n^3)\right) $. In the case of F-WMMSE utilizing the RLS-SOMP algorithm, with the condition that $N_t \geq KD$, the overall complexity can be represented as  $\mathcal{O}\left(\mathcal{I}N_tG_tK^2D^2+\mathcal{I} \sum_{n=1}^{N_t}(n^2KD+n^3)\right) $. When compared to WMMSE, it is clear that F-WMMSE combined with the RLS-SOMP algorithm necessitates a higher-order complexity.

\subsection{MIL-based Fast Implementation}
To enhance the efficiency of F-WMMSE, employing the atom matching process is simplified in the expense of memory storage and the MIL is suggested to reduce the complexity associated with RLS-SOMP.
First, the atom matching process requires the inner product of the sensing matrix and the residual, i.e., $\underset{g\in\bm{\Gamma}}{{\rm arg \ max}} \ \left\Vert   [\mathbf{D}]_{:,g} ^{H} {\mathbf{R}}  \right\Vert_2^2$. 

 The matching expression is given by 
\begin{equation}\label{DR}
 [\mathbf{D}]_{:,g} ^{H} {\mathbf{R}} =[\mathbf{D}]_{:,g} ^{H}\mathbf{Y}-[\mathbf{D}]_{:,g} ^{H}[\mathbf{D}]_{:,\bm{\Lambda}}\mathbf{X}^\star.
\end{equation}
We can observe that the matching process can be simplified by pre-calculation and storage of $[\mathbf{D}]_{:,g} ^{H}\mathbf{Y}$ and $[\mathbf{D}]_{:,g_1}^H[\mathbf{D}]_{:,g_2}$, $\forall g,g_1,g_2$. Hence, the matching complexity in the $n$-th iteration is decreased into $\mathcal{O}(nGKD)$ by calculating $[\mathbf{D}]_{:,g} ^{H}[\mathbf{D}]_{:,\bm{\Lambda}}\mathbf{X}^\star,\forall g$.

Denoted by $\bm{\Lambda}_n$ and $g_n^\star$ the support and the selected atom in the $n$-th iteration. 
\begin{equation}\label{DDI}
	\begin{aligned}
	&	\left([\mathbf{D}]^H_{:,\bm{\Lambda}_{n}}[\mathbf{D}] _{:,\bm{\Lambda}_{n}}+\zeta\mathbf{I}_n\right)^{-1} \\
	&	= \begin{bmatrix}
			[\mathbf{D}]^H_{:,\bm{\Lambda}_{n-1}}[\mathbf{D}] _{:,\bm{\Lambda}_{n-1}}+\zeta\mathbf{I}_{n-1} & [\mathbf{D}] _{:,\bm{\Lambda}_{n-1}}^H[\mathbf{D}]_{:,g^\star_n} \\ [\mathbf{D}]^H_{:,g^\star_n}[\mathbf{D}] _{:,\bm{\Lambda}_{n-1}} & [\mathbf{D}]^H_{:,g^\star_n}[\mathbf{D}] _{:,g^\star_n}+\zeta
		\end{bmatrix}^{-1} \\
		& =  \begin{bmatrix}
			\left([\mathbf{D}]^H_{:,\bm{\Lambda}_{n-1}}[\mathbf{D}] _{:,\bm{\Lambda}_{n-1}}+\zeta\mathbf{I}_n \right)^{-1}+\eta_n\mathbf{v}_n\mathbf{v}_n^H & -\eta_n\mathbf{v}_{n} \\ -\eta_n\mathbf{v}_n^H &\eta_n
		\end{bmatrix},
	\end{aligned}
\end{equation}
 where $\mathbf{v}_n\triangleq\mathbf{U}_{n-1}^H[\mathbf{D}] _{:,g^\star_n}$ with $\mathbf{U}_{n-1}\triangleq [\mathbf{D}]_{:,\bm{\Lambda}_{n-1}} \left([\mathbf{D}]^H_{:,\bm{\Lambda}_{n-1}}[\mathbf{D}] _{:,\bm{\Lambda}_{n-1}}+\zeta\mathbf{I}_{n-1}\right)^{-1}$, and
 $\eta_n\triangleq \frac{1}{\Vert[\mathbf{D}] _{:,g^\star_n}\Vert^2_2+\zeta-\Vert [\mathbf{D}]^H_{:,\bm{\Lambda}_{n-1}} \mathbf{v}_n\Vert_2^2}$.
 
 Consider the recursion of $\mathbf{U}_n$
 \begin{equation}
 	\begin{aligned}
 		\mathbf{U}_n=&\begin{bmatrix}
 			[\mathbf{D}]_{:,\bm{\Lambda}_{n-1}} & [\mathbf{D}] _{:,g^\star_n}
 		\end{bmatrix}\left([\mathbf{D}]^H_{:,\bm{\Lambda}_{n}}[\mathbf{D}] _{:,\bm{\Lambda}_{n}}+\zeta\mathbf{I}_n\right)^{-1} \\
 		=& \begin{bmatrix}
 			\mathbf{U}_{n-1}+ \eta_n\mathbf{t}_n\mathbf{v}_n^H & \eta_n\mathbf{t}_n
 		\end{bmatrix},
 	\end{aligned}
 \end{equation}
 where $\mathbf{t}_n\triangleq	[\mathbf{D}]_{:,\bm{\Lambda}_{n-1}} \mathbf{v}_n-[\mathbf{D}]_{:,g^\star_n}$.
 
 \begin{equation}
 	\begin{aligned}
 		\mathbf{X}^\star_n&=\mathbf{U}_n^H\mathbf{Y} \\
 		&=\begin{bmatrix}
 		\mathbf{U}_{n-1}\mathbf{Y}+\eta_n\mathbf{v}_n\mathbf{t}_n^H\mathbf{Y} \\ \eta_n\mathbf{t}_n^H\mathbf{Y}
 		\end{bmatrix} \\&
 		=\begin{bmatrix}
 		 \mathbf{X}^\star_{n-1} \\ 0
 		\end{bmatrix}
 	+\begin{bmatrix}
 		\mathbf{v}_n \\ 1
 	\end{bmatrix}\eta_n\mathbf{t}_n^H\mathbf{Y}.
 	\end{aligned}
 \end{equation}
 
 Recalling the residual updating process in line 7 of Algorithm \ref{RLS-SOMP}, and considering the $n$-th iteration for $[\mathbf{D}]_{:,\bm{\Lambda}_n}\mathbf{X}_n^\star$ calculation:
 \begin{equation}
 	\begin{aligned}
 	&[\mathbf{D}]_{:,\bm{\Lambda}_n}\left(  \begin{bmatrix}
 			\mathbf{X}^\star_{n-1} \\ 0
 		\end{bmatrix}
 		+\begin{bmatrix}
 			\mathbf{v}_n \\ 1
 		\end{bmatrix}\eta_n\mathbf{t}_n^H\mathbf{Y}\right) \\
 	&	=	[\mathbf{D}]_{:,\bm{\Lambda}_{n-1}} \mathbf{X}^\star_{n-1} +
 		[\mathbf{D}]_{:,\bm{\Lambda}_{n-1}}\mathbf{v}_n\mathbf{t}_n^H\mathbf{Y}\\&\ \ \ + [\mathbf{D}]_{:,g^\star_n} \mathbf{t}_n^H\mathbf{Y}. 
 	\end{aligned}
 \end{equation} 
 
 The above recursion equation identifies that the first term can be obtained by the last iteration. Moreover, we can derive $\mathbf{t}_n^H\mathbf{Y}$ as
 \begin{equation}
 	\mathbf{t}_n^H\mathbf{Y}=\mathbf{v}_n^H[\mathbf{D}]_{:,\bm{\Lambda}_{n-1}}^H\mathbf{Y} -[\mathbf{D}]_{:,g^\star_n}^H\mathbf{Y}.
 \end{equation}
 As $\mathbf{D}^H\mathbf{Y}$ can be calculated only once and storage,
 the above equation only requires a complexity of $\mathcal{O}(nKD)$ for vector-matrix multiplication. Moreover, calculating $\mathbf{v}_n$ and $[\mathbf{D}]_{:,\bm{\Lambda}_{n-1}}\mathbf{v}_n$ require a complexity of $\mathcal{O}(nKD)$.
  Hence, the total complexity for calculating coefficient and residual is $\mathcal{O}(nKD)$.
  It also indicates that the overall complexity of fast RLS-SOMP is $\mathcal{O}(nGKD)$ induced by the atom matching process. Furthermore, the complexity of F-WMMSE with fast RLS-SOMP can be given by $\mathcal{O}\left(\mathcal{I}\sum_{n=1}^{N_t} nGKD\right)$.
  
\section{Simulation Results}\label{SR}

The system operates at a central frequency of $3$ GHz. The BS is equipped with $N_t=16$ MAs, serving $K=4$ users equipped with $N_r=4$ MAs with $D=4$ data streams. The noise power is normalized to $1$ such that the signal-to-noise ratio (SNR) is expressed by $P$. The user and scatterer locations are uniformly distributed with azimuth angles $\phi_{k,l}$ and elevation angles $\theta_{k,l}$ ranging within $[-1,1]$ for all $k,l$. All users are assumed to have an identical number of channel' paths.  All users are fair such that the priority $\alpha_k$ is set to $1$ for all users.
The evaluated methods include MMSE, WMMSE, and the proposed F-WMMSE. For the MMSE and WMMSE methods, the antenna positions are fixed and arranged in a uniform planar array (UPA) with half-wavelength spacing. In contrast, for the F-WMMSE method, the transmit and receive antenna positions are assumed to be adjustable within allowed transmit and receive movable regions, as defined by square regions in Eqs. (\ref{Ut}) and (\ref{Ur}). The dimensions of these transmit and receive regions are determined by $U_t$ and $U_r$, respectively.

 \begin{figure}
	\centering
	\subfigure[SNR = $-5$ dB.]{
		\includegraphics[width=3.1in]{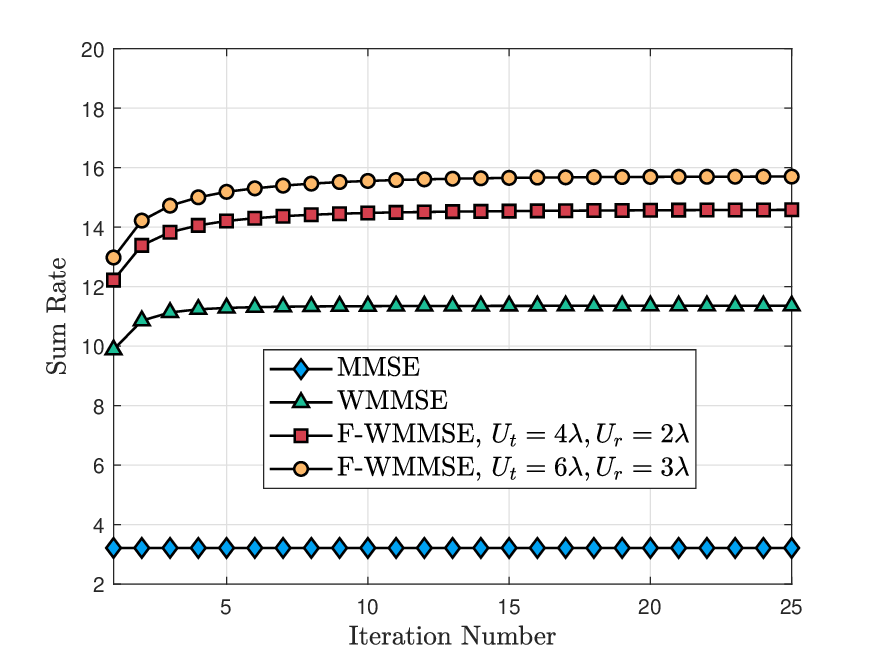}
	}
	\\    
	\subfigure[SNR = $5$ dB.]{
		\includegraphics[width=3.1in]{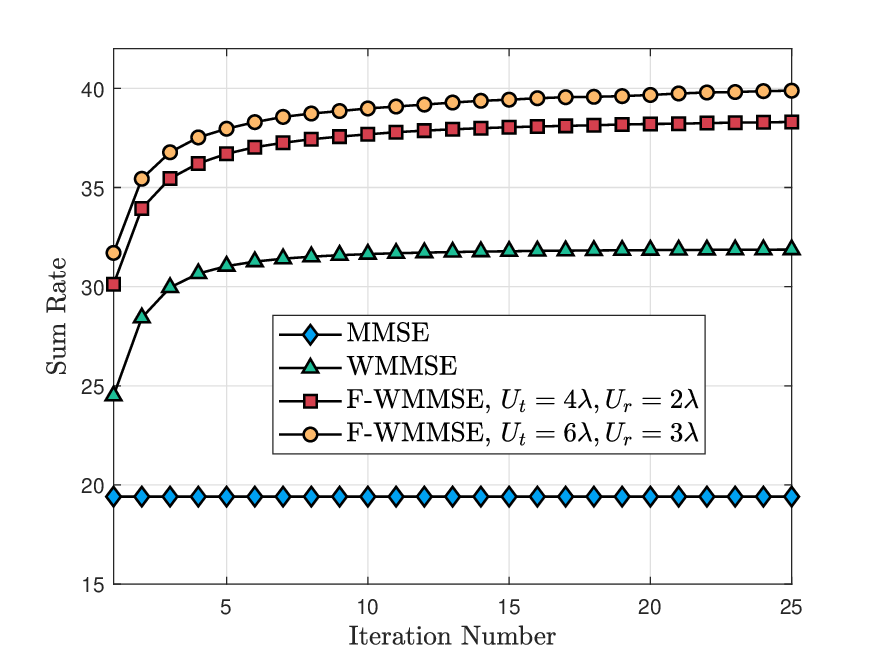}
	} 
	\caption{The sum rate versus the iteration number.}
	\label{sr_iter}
\end{figure}

The first simulation evaluates the convergence of the proposed F-WMMSE method in terms of sum rate. The number of iterations ranges from $1$ to $25$, with the number of channel paths $L$ set to $10$. Two SNR levels $-5$ and $5$ dB are considered, along with two movable region settings  $(U_t,U_r)=\{(4\lambda,2\lambda),(6\lambda,3\lambda)\}$. For example,  $(U_t,U_r)=(4\lambda,2\lambda)$ constrains the transmit and receive antennas to movement within square regions of $4\lambda\times 4\lambda$ and $2\lambda \times 2\lambda$, respectively. As shown in Fig. \ref{sr_iter}, the proposed F-WMMSE exhibits convergence as the iteration count $\mathcal{I}$ increases, reaching a stable state after relatively few iterations, similar to WMMSE. Notably, the F-WMMSE method achieves a significant performance improvement over WMMSE. Additionally, as the movable region expands, the sum rate of F-WMMSE further increases.

\begin{figure}
	\centering
	\subfigure[$K=2$.]{
		\includegraphics[width=3.1in]{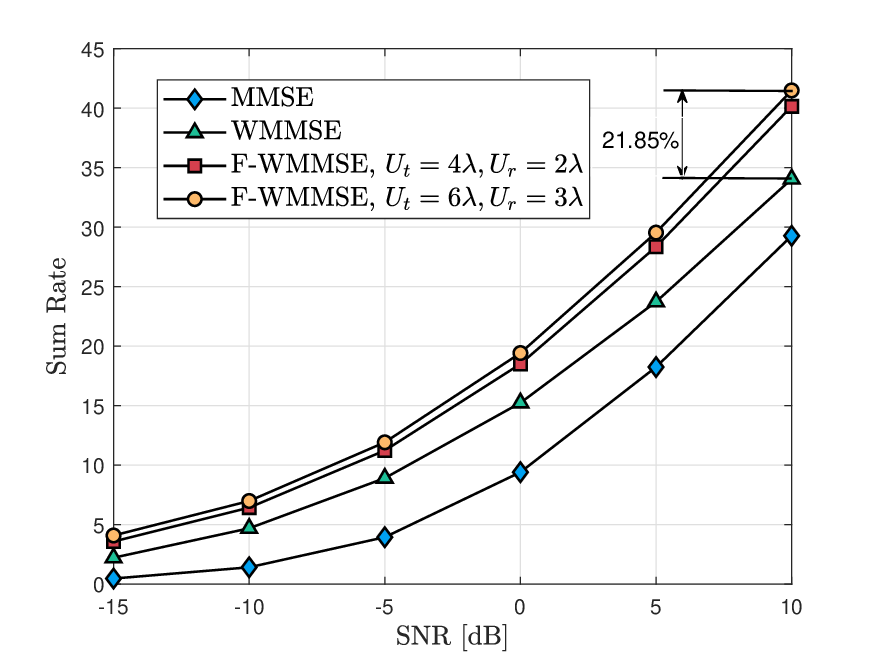}
	}
	\\    
	\subfigure[$K=4$.]{
		\includegraphics[width=3.1in]{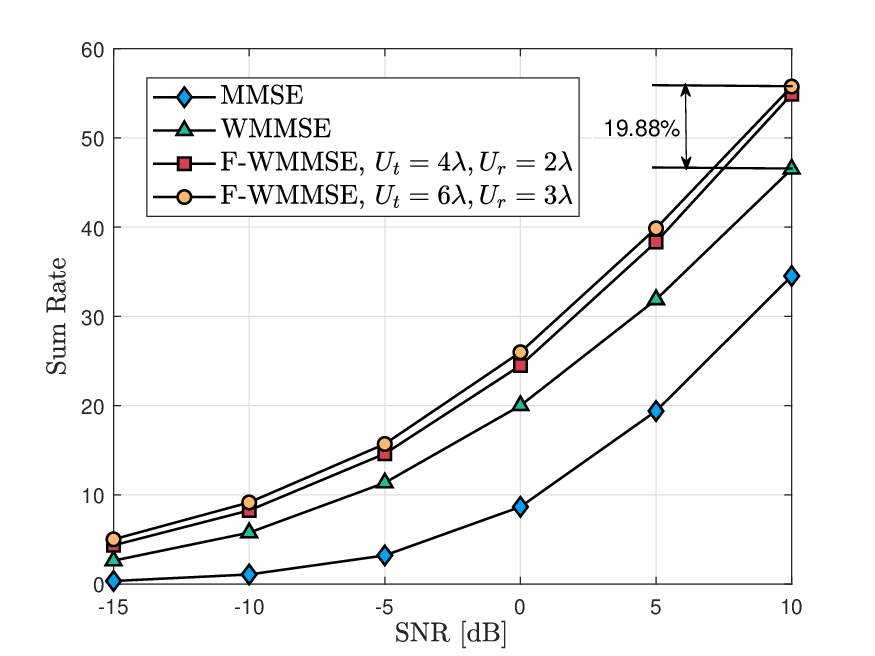}
	} 
	\caption{The sum rate versus SNR.}
	\label{sr_snr}
\end{figure}

  \begin{figure}
	\centering 
	\includegraphics[width=3.1in]{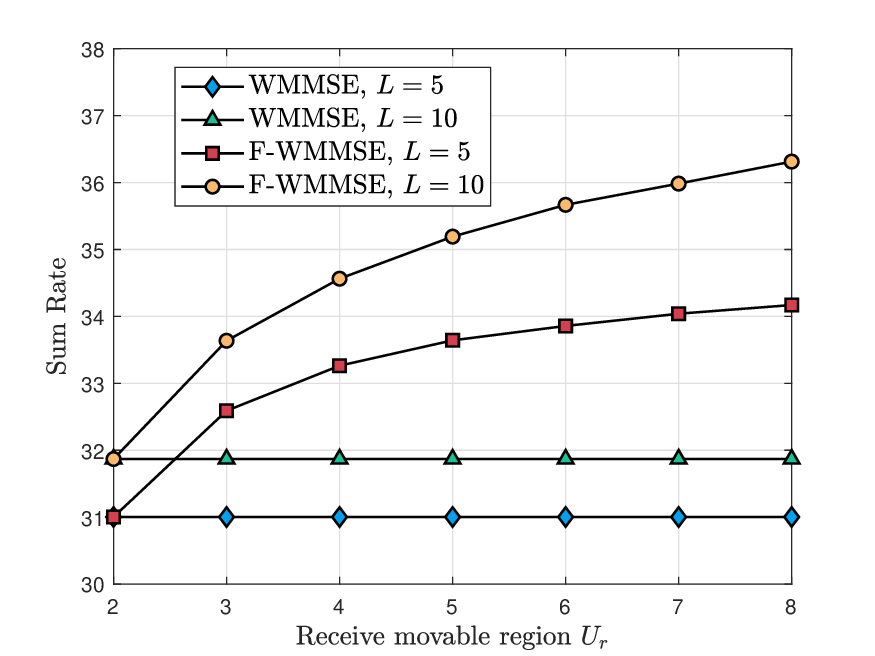}
	\caption{The sum rate versus the receive movable region $U_r$.}\label{RM} 
\end{figure} 

 \begin{figure}
	\centering 
	\includegraphics[width=3.1in]{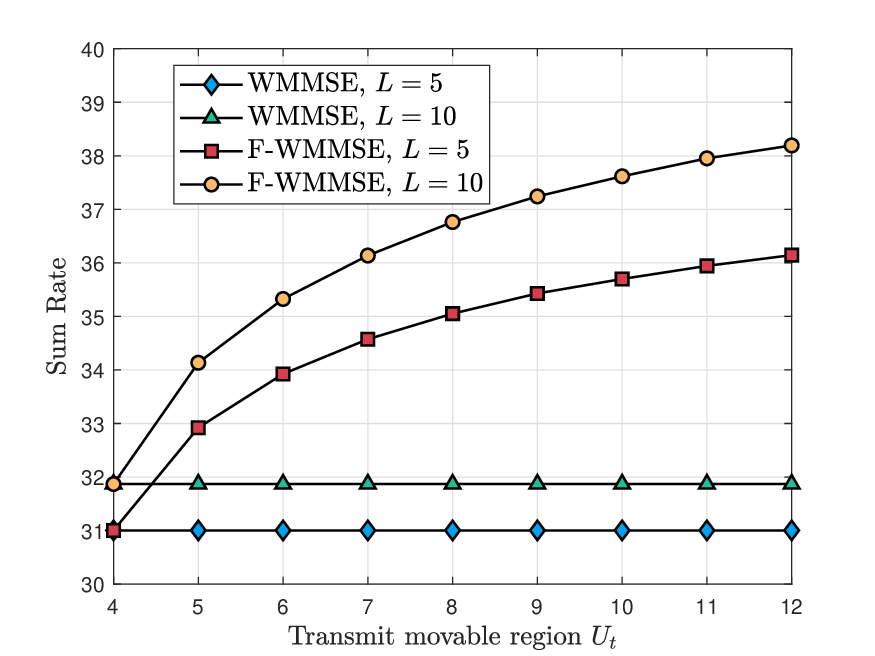}
	\caption{The sum rate versus the transmit movable region $U_t$.}\label{TM} 
\end{figure} 

  \begin{figure}
 	\centering 
 	\includegraphics[width=3.1in]{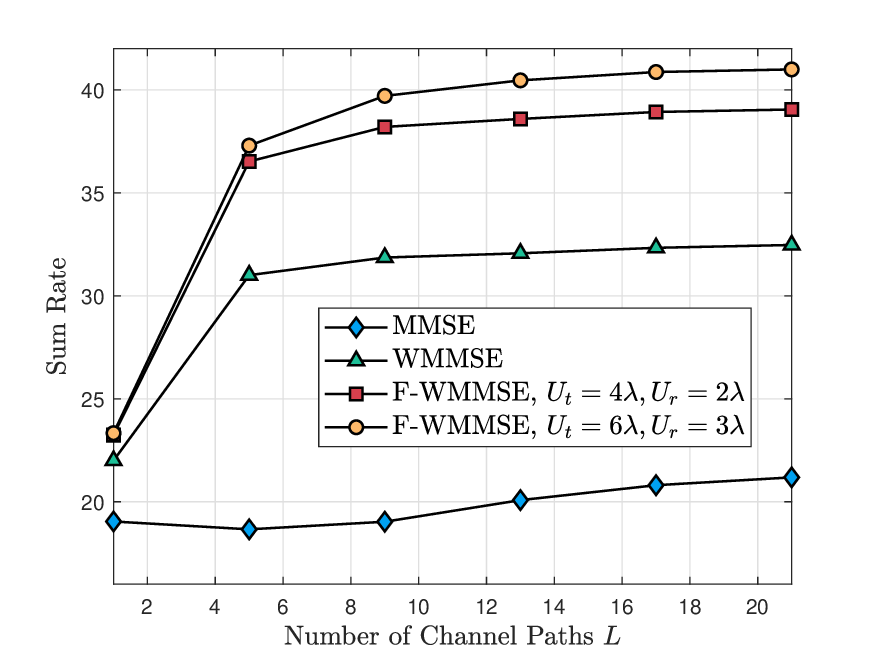}
 	\caption{The sum rate versus the number of channel paths $L$.}\label{path} 
 \end{figure}

The second simulation evaluates the impact of SNR on the proposed F-WMMSE method, with SNR ranging from $-15$ to $10$ dB, and the number of channel paths, $L$, set to $10$. Similar to the first simulation, two movable region configurations are considered:  $(U_t,U_r)=\{(4\lambda,2\lambda),(6\lambda,3\lambda)\}$. The results show that the sum rate increases with SNR across all methods, with MMSE exhibiting the smallest growth trend and F-WMMSE the largest. Notably, F-WMMSE achieves a significant performance improvement over WMMSE.  For example, F-WMMSE with $(U_t,U_r)=(6\lambda,3\lambda)$ has a performance gain of $21.85\%$ over WMMSE at SNR $=10$ dB when the number of users is $K=2$, though this gain decreases to $19.88\%$ when $K=4$.  Furthermore, the smaller movable region $(U_t,U_r)=(4\lambda,2\lambda)$ yields a lower, yet still substantial, performance gain.

The third simulation assesses the impact of the movable region on the proposed F-WMMSE method, including both the transmit and receive movable regions. We evaluate these regions separately by fixing one at half-wavelength spacing and varying the other. As shown in Fig. \ref{RM}, with SNR set to $5$ dB, the number of channel paths set to $L\in\{5,10\}$, and the receive movable region $U_r$ ranging from $\lambda$ to $4\lambda$ with steps of $0.5\lambda$ for all users. Notably, to isolate the impact of the receive movable region, the transmit antenna region is fixed as a $4\times 4$ UPA. The results indicate that the sum rate of F-WMMSE increases with $U_r$, although the rate of increase diminishes as $U_r$ grows.  Notably, $U_r=\lambda$ corresponds to antennas arranged in a fixed $2\times 2$ UPA, indicating no antenna position adjustment. Moreover, F-WMMSE with $L = 10$ shows a significant performance gain over $L=5$, highlighting the impact of the number of paths on MAs. 
Fig. \ref{TM} illustrates the influence of the transmit movable region $U_t$ on the F-WMMSE sum rate. Here, parameters mirror those in Fig. \ref{RM}, except the receive antenna region is fixed at a $2 \times 2$  UPA, while $U_t$ ranges from $2\lambda$ to $6\lambda$ in steps of $0.5\lambda$. The results similarly show that the F-WMMSE sum rate increases with $U_t$, underscoring the importance of the transmit movable region. As $U_t$ grows, the sum rate gain decreases, with this trend significantly affected by the number of channel paths, $L$. Figs. \ref{RM} and \ref{TM} demonstrate that increasing both the transmit and receive movable regions can enhance the sum rate of the proposed F-WMMSE method.

Finally, we evaluate the impact of the number of channel paths on the proposed F-WMMSE, as shown in Fig. \ref{path}. In this scenario, the SNR is set to $5$ dB, the two movable region configurations are $(U_t,U_r)=\{(4\lambda,2\lambda),(6\lambda,3\lambda)\}$, and the number of channel paths $L$ ranges from $1$ to $21$ in steps of $4$.
The results show that as $L$ increases,both WMMSE and F-WMMSE exhibit performance improvements, with F-WMMSE showing a more pronounced enhancement. For example, as $L=1$, F-WMMSE and WMMSE achieve comparable sum rates. However, as $L$ reaches $3$, the sum rate difference between the two methods becomes significant, and this performance gap continues to widen with increasing $L$, although it starts to converge from  $L=13$. Additionally, F-WMMSE with $(U_t,U_r)=(6\lambda,3\lambda)$ shows a performance advantage over $(U_t,U_r)=(4\lambda,2\lambda)$ as $L$ increases. This suggests that a larger movable region is necessary to fully exploit the potential benefits in a relatively rich scattering environment.

\section{Conclusions}\label{Con} 

This paper explored MA-enabled MU-MIMO communications to enhance the sum rate through joint optimization of antenna positioning and beamforming. The WMMSE algorithm addressed a matrix-weighted sum-MSE minimization problem, equivalent to WSR maximization through the introduction of a weight matrix, thereby enabling exceptional performance in joint precoder and combiner optimization for MU-MIMO. In this context, we incorporated antenna position optimization within the WMMSE framework. To this end, we reformulated the WMMSE subproblems using RLS and establish a sparse optimization framework that integrates both antenna positioning and beamforming optimization. This complete procedure, termed F-WMMSE, has a higher time complexity than traditional WMMSE; however, this complexity is mitigated with our proposed acceleration technique. Simulations demonstrated that F-WMMSE achieved a notable improvement in sum rate compared to traditional WMMSE, with performance gains that increase with larger movable regions and a greater number of channel paths.

\begin{appendices}

\section{Proof of Proposition \ref{pro_1} }\label{W_bian} 
	First, we rewrite the equivalent MSE matrix as
\begin{equation}
	\begin{aligned}
		&	\widetilde{\mathbf{E}}_k= \left(\mathbf{I}_D- \mathbf{W}_k^H\mathbf{H}_k\mathbf{F}_k
		\right)\left(\mathbf{I}_D- \mathbf{W}_k^H\mathbf{H}_k\mathbf{F}_k
		\right)^H \\
		&+\mathbf{0}_{D\times D}+\mathbf{W}_k^H\mathbf{H}_k\mathbf{F}_{\lnot k}\mathbf{F}_{\lnot k}^H\mathbf{H}_k^H\mathbf{W}_k+  \frac{\sigma^2_k}{P} {\rm Tr}\left(
		\mathbf{F}\mathbf{F}^H\right)
		\mathbf{W}^H_k\mathbf{W}_k,
	\end{aligned}
\end{equation}
where $\mathbf{F}_{\lnot k}\triangleq [\mathbf{F}_1,\cdots, \mathbf{F}_{k-1},\mathbf{F}_{k+1},\cdots,\mathbf{F}_K]\in\mathbb{C}^{N_t\times (K-1)D}$ represents the precoder for all users' data streams except for user $k$.

Then, its trace is given by
\begin{equation}\label{ETR}
	\begin{aligned}
		&{\rm Tr}\left(\widetilde{ \mathbf{E}}_k\right) \\ =& \left\Vert \mathbf{I}_D-\mathbf{W}_k^H\mathbf{H}_k\mathbf{F}_k \right\Vert_F^2 +\left\Vert \mathbf{0}_{D\times (K-1)D}-\mathbf{W}_k^H\mathbf{H}_k\mathbf{F}_{\lnot k}\right\Vert_F^2 \\
		&+   \frac{\sigma^2_k}{P} {\rm Tr}\left(
		\mathbf{F}\mathbf{F}^H\right)
		{\rm Tr}\left(\mathbf{W}^H_k\mathbf{W}_k \right)\\
		=& \left\Vert \left[ \mathbf{I}_{D} \ \mathbf{0}_{D\times (K-1)D}\right]-\left[ \mathbf{W}_k^H\mathbf{H}_k\mathbf{F}_k \ \mathbf{W}_k^H\mathbf{H}_k\mathbf{F}_{\lnot k}
		\right] \right\Vert_F^2 \\ & +\frac{\sigma^2_k}{P} {\rm Tr}\left(
		\mathbf{F}\mathbf{F}^H\right)
		\left\Vert\mathbf{W}_k\right\Vert_F^2\\
		=& \left\Vert \left[\mathbf{I}_{D} \ \mathbf{0}_{D\times (K-1)D}\right]- \mathbf{W}_k^H\mathbf{H}_k\left[ \mathbf{F}_k \ \mathbf{F}_{\lnot k}
		\right]
		\right\Vert_F^2 \\ & +\frac{\sigma^2_k}{P} {\rm Tr}\left(
		\mathbf{F}\mathbf{F}^H\right)
		\left\Vert\mathbf{W}_k\right\Vert_F^2.
	\end{aligned}
\end{equation}

To make $\mathbf{F}=\left[ \mathbf{F}_k \ \mathbf{F}_{\lnot k}
\right]$, we need to place $\mathbf{F}_k$ into the appropriate position in $\mathbf{F}_{\lnot k}$. Correspondingly, 
$\left[\mathbf{I}_{D} \ \mathbf{0}_{D\times (K-1)D}\right]$ requires change for consistency. Thus, the first term in Eq. (\ref{ETR}) can be simplified by $\left\Vert\begin{bmatrix}
	\mathbf{0}_{D\times (k-1)D}, \mathbf{I}_D, \mathbf{0}_{D\times (K-k)D}
\end{bmatrix}- \mathbf{W}_k^H\mathbf{H}_k\mathbf{F}
\right\Vert_F^2$.
We define $\widetilde{\mathbf{I}}_k\triangleq
\begin{bmatrix}
	\mathbf{0}_{D\times (k-1)D}, \mathbf{I}_D, \mathbf{0}_{D\times (K-k)D}
\end{bmatrix}^T\in\mathbb{C}^{KD\times K}$ and obtain
\begin{equation}
	{\rm Tr}\left(\widetilde{\mathbf{E}}_k
	\right)=	\left\Vert \widetilde{\mathbf{I}}_k- \mathbf{F}^H\mathbf{H}_k^H\mathbf{W}_k
	\right\Vert_F^2  +\frac{\sigma^2_k}{P} {\rm Tr}\left(
	\mathbf{F}\mathbf{F}^H\right)
	\left\Vert\mathbf{W}_k\right\Vert_F^2.
\end{equation}

Hence, problem (\ref{FW1}) can be formulated and its solution can be easily obtained by Eq. (\ref{FW1_S}).

	\section{Proof of Proposition \ref{RLSSOMP} }\label{GP} 
	The objective of (\ref{CSS}) is different from the standard SOMP due to the regularized term. To solve the nonlinear sparse recovery problem, the greedy pursuit (GP) criterion \cite{GP} is employed. In the $n$-th iteration, GP selects the atom  with the maximum gradient component and update the variable:
		
1) GP selects a column index in $\mathbf{D}$ and update the support set 	$\bm{\Lambda}=\bm{\Lambda} \cup g^\star$: 
	\begin{equation}\label{gst}
	g^\star=\underset{g}{\rm arg \ max} \ \left\Vert \nabla _{\mathbf{X}_g}^{(n)}\right\Vert_2^2,
	\end{equation}
	where $\nabla _{\mathbf{X}_g}^{(n)}$ is the $g$-th row of the gradient $\nabla _{\mathbf{X}}^{(n)}$ of the objective function, provide by	\begin{equation}\label{DXn}
		\nabla _{\mathbf{X}}^{(n)}=2\mathbf{D}^H(\mathbf{D}\mathbf{X}^{(n-1)}-\mathbf{Y})+2\zeta \mathbf{X}^{(n-1)}.
	\end{equation}
	
2) $\mathbf{X}^{(n)}$ is updated by maximizing the objective given $\bm{\Lambda}$:
\begin{equation}
	\mathbf{X}^{(n)}=\left([\mathbf{D}]_{:,\bm{\Lambda}}^{H}[\mathbf{D}]_{:,\bm{\Lambda}}+\zeta \mathbf{I}_n\right)^{-1}[\mathbf{D}]_{:,\bm{\Lambda}}^{H}\mathbf{Y}.
\end{equation}

Observing that $2\mathbf{D}^H(\mathbf{D}\mathbf{X}^{(n-1)}-\mathbf{Y})=-2\mathbf{D}^H\mathbf{R}^{(n-1)}$ where $\mathbf{R}^{(n-1)}\triangleq \mathbf{Y}-\mathbf{D}\mathbf{X}^{(n-1)}$ represents the residual signal in SOMP, and noting that the second term in (\ref{DXn}) does not affect (\ref{gst}). It identifies that, despite the inclusion of the regularization term, the atom selection criterion remains consistent with that of SOMP. The key distinction lies in the replacement of the LS calculation with RLS. Consequently, we designate this algorithm as RLS-SOMP. 
\section{Proof of Proposition \ref{F_pro} }\label{F_bian} 
First, we simplify the objective of WMMSE precoder problem as
\begin{equation}\label{BE}
	\begin{aligned}
		&	\sum_{k=1}^{K}	\alpha_k{\rm Tr}\left(\mathbf{B}_k\widetilde{\mathbf{E}}_k\right)=	{\rm Tr}\left( \sum_{k}	\alpha_k\mathbf{B}_k  \mathbf{W}^H_k \mathbf{H}_k\mathbf{F}\mathbf{F}^H\mathbf{H}_k^H \mathbf{W}_k\right)   \\&  \ 
		- {\rm Tr}\left( \sum_{k}	\alpha_k\mathbf{B}_k \mathbf{F}_k^H\mathbf{H}_k^H\mathbf{W}_k \right) -{\rm Tr}\left( \sum_{k}	\alpha_k\mathbf{B}_k\mathbf{W}_k^H\mathbf{H}_k\mathbf{F}_k\right)\\ &  \ +{\rm Tr}\left(\sum_{k}\alpha_k \mathbf{B}_k \right)
		+ {\rm Tr}\left(\sum_{k} \frac{	\alpha_k\sigma^2_k}{P} {\rm Tr}\left(
		\mathbf{F}\mathbf{F}^H\right)\mathbf{B}_k\mathbf{W}^H_k\mathbf{W}_k\right).
	\end{aligned}
\end{equation}

Denoted by $\widetilde{\mathbf{H}}_k\triangleq \sqrt{\alpha_k}\mathbf{B}^{\frac{1}{2}}_k \mathbf{W}_k^H\mathbf{H}_k\in\mathbb{C}^{D\times N_t}$, $\widetilde{\mathbf{H}}\triangleq \left[ \widetilde{\mathbf{H}}_1^T,\cdots, \widetilde{\mathbf{H}}_K^T
\right]^T\in\mathbb{C}^{KD\times N_t}
$ and $\widetilde{\mathbf{B}}\triangleq {\rm blkdiag}\left( \alpha_1\mathbf{B}_1,\cdots,\alpha_K\mathbf{B}_K
\right)\in\mathbb{C}^{KD\times KD}$, noticing $\mathbf{B}_k \succeq 0$ such that $\mathbf{B}_k = \mathbf{B}_k^H$,
then those terms in Eq. (\ref{BE}) can be further simplified, respectively:
\begin{equation}
	\begin{aligned}
		&{\rm Tr}\left( \sum_{k}	\alpha_k\mathbf{B}_k  \mathbf{W}^H_k \mathbf{H}_k\mathbf{F}\mathbf{F}^H\mathbf{H}_k^H \mathbf{W}_k\right)    \\= &
		{\rm Tr}\left( \sum_{k}\mathbf{F}^H\mathbf{H}_k^H \mathbf{W}_k	\alpha_k\mathbf{B}_k  \mathbf{W}^H_k \mathbf{H}_k\mathbf{F}\right) \\
		=&  	{\rm Tr}\left(
		\sum_{k} \mathbf{F}^H \widetilde{\mathbf{H}}_k^H \widetilde{\mathbf{H}}_k\mathbf{F}
		\right) \\
		=& {\rm Tr}\left( \mathbf{F}^H \widetilde{\mathbf{H}}^H\widetilde{\mathbf{H}}{\mathbf{F}}
		\right),
	\end{aligned}
\end{equation}
\begin{equation}
	\begin{aligned}
		{\rm Tr}\left( \sum_{k}	\alpha_k\mathbf{B}_k \mathbf{F}_k^H\mathbf{H}_k^H\mathbf{W}_k \right)= &
		{\rm Tr}\left( 
		\sum_{k} \sqrt{\alpha_k}\mathbf{F}_k^H \widetilde{\mathbf{H}}_k^H \mathbf{B}_k^{\frac{1}{2}}
		\right) \\
		=& {\rm Tr}\left(\widetilde{\mathbf{H}}^H \widetilde{\mathbf{B}}^{\frac{1}{2}} \mathbf{F}^H \right),
	\end{aligned}
\end{equation}
\begin{equation}
	\begin{aligned}
		{\rm Tr}\left( \sum_{k}	\alpha_k\mathbf{B}_k\mathbf{W}_k^H\mathbf{H}_k\mathbf{F}_k\right)= &
		{\rm Tr}\left( \sum_{k}	\sqrt{\alpha_k}\mathbf{B}_k^{\frac{1}{2}}\widetilde{\mathbf{H}}_k\mathbf{F}_k\right)\\
		=&  {\rm Tr}\left(\mathbf{F} \widetilde{\mathbf{B}}^{\frac{1}{2}}\widetilde{\mathbf{H}} \right),
	\end{aligned}
\end{equation}
\begin{equation}\label{BEN}
	{\rm Tr}\left( \sum_{k}\alpha_k\mathbf{B}_k
	\right)= {\rm Tr}\left( \widetilde{\mathbf{B}}
	\right).
\end{equation} 

Combining Eqs. (\ref{BE})-(\ref{BEN}), we can obtain
\begin{equation}
	\begin{aligned}
		&	\sum_{k=1}^{K} \alpha_k {\rm Tr}\left(
		\mathbf{B}_k\widetilde{\mathbf{E}}_k\right) \\ &   
		= \left\Vert 
		\widetilde{\mathbf{B}}^{\frac{1}{2}} -\widetilde{\mathbf{H}}\mathbf{F}
		\right\Vert_F^2	+  \sum_{k} \frac{	\alpha_k\sigma^2_k}{P} \left\Vert
		\mathbf{F}\right\Vert_F^2\left\Vert  \mathbf{W}_k\mathbf{B}_k^\frac{1}{2} \right\Vert_F^2.
	\end{aligned}
\end{equation}

Therefore, problem (\ref{FW-F1}) and its solution (\ref{FW-F2}) can be derived.
\end{appendices}

\bibliographystyle{IEEEtran}
\bibliography{reference.bib}

\begin{thebibliography}{10}
\providecommand{\url}[1]{#1}
\csname url@samestyle\endcsname
\providecommand{\newblock}{\relax}
\providecommand{\bibinfo}[2]{#2}
\providecommand{\BIBentrySTDinterwordspacing}{\spaceskip=0pt\relax}
\providecommand{\BIBentryALTinterwordstretchfactor}{4}
\providecommand{\BIBentryALTinterwordspacing}{\spaceskip=\fontdimen2\font plus
\BIBentryALTinterwordstretchfactor\fontdimen3\font minus
  \fontdimen4\font\relax}
\providecommand{\BIBforeignlanguage}[2]{{%
\expandafter\ifx\csname l@#1\endcsname\relax
\typeout{** WARNING: IEEEtran.bst: No hyphenation pattern has been}%
\typeout{** loaded for the language `#1'. Using the pattern for}%
\typeout{** the default language instead.}%
\else
\language=\csname l@#1\endcsname
\fi
#2}}
\providecommand{\BIBdecl}{\relax}
\BIBdecl

\bibitem{mmw1}
R.~He, B.~Ai, G.~Wang, M.~Yang, C.~Huang, and Z.~Zhong, ``Wireless channel
  sparsity: Measurement, analysis, and exploitation in estimation,'' \emph{IEEE
  Wireless Communications}, vol.~28, no.~4, pp. 113--119, 2021.

\bibitem{mmw2}
S.~Yang, C.~Xie, D.~Wang, and Z.~Zhang, ``Fast multibeam training for mmwave
  mimo systems with subconnected hybrid beamforming architecture,'' \emph{IEEE
  Systems Journal}, vol.~17, no.~2, pp. 2939--2949, 2023.

\bibitem{BN1}
B.~Ning, Z.~Tian, W.~Mei, Z.~Chen, C.~Han, S.~Li, J.~Yuan, and R.~Zhang,
  ``Beamforming technologies for ultra-massive mimo in terahertz
  communications,'' \emph{IEEE Open Journal of the Communications Society},
  vol.~4, pp. 614--658, 2023.

\bibitem{RIS1}
C.~Huang, A.~Zappone, G.~C. Alexandropoulos, M.~Debbah, and C.~Yuen,
  ``Reconfigurable intelligent surfaces for energy efficiency in wireless
  communication,'' \emph{IEEE Transactions on Wireless Communications},
  vol.~18, no.~8, pp. 4157--4170, 2019.

\bibitem{RIS2}
Q.~Wu and R.~Zhang, ``Intelligent reflecting surface enhanced wireless network
  via joint active and passive beamforming,'' \emph{IEEE Transactions on
  Wireless Communications}, vol.~18, no.~11, pp. 5394--5409, 2019.

\bibitem{BN2}
B.~Ning, Z.~Chen, W.~Chen, Y.~Du, and J.~Fang, ``Terahertz multi-user massive
  mimo with intelligent reflecting surface: Beam training and hybrid
  beamforming,'' \emph{IEEE Transactions on Vehicular Technology}, vol.~70,
  no.~2, pp. 1376--1393, 2021.

\bibitem{NF1}
H.~Lu and Y.~Zeng, ``Near-field modeling and performance analysis for
  multi-user extremely large-scale mimo communication,'' \emph{IEEE
  Communications Letters}, vol.~26, no.~2, pp. 277--281, 2022.

\bibitem{NF2}
S.~Yang, W.~Lyu, Z.~Hu, Z.~Zhang, and C.~Yuen, ``Channel estimation for
  near-field xl-ris-aided mmwave hybrid beamforming architectures,'' \emph{IEEE
  Transactions on Vehicular Technology}, vol.~72, no.~8, pp. 11\,029--11\,034,
  2023.

\bibitem{Liquid}
Y.~Huang, L.~Xing, C.~Song, S.~Wang, and F.~Elhouni, ``Liquid antennas: Past,
  present and future,'' \emph{IEEE Open Journal of Antennas and Propagation},
  vol.~2, pp. 473--487, 2021.

\bibitem{MEMS}
Z.~Baghchehsaraei, ``Waveguide-integrated mems concepts for tunable
  millimeter-wave systems,'' Ph.D. dissertation, KTH Royal Institute of
  Technology, 2014.

\bibitem{Pixel}
S.~Song and R.~D. Murch, ``An efficient approach for optimizing frequency
  reconfigurable pixel antennas using genetic algorithms,'' \emph{IEEE
  Transactions on Antennas and Propagation}, vol.~62, no.~2, pp. 609--620,
  2014.

\bibitem{Stepper}
S.~Basbug, ``Design and synthesis of antenna array with movable elements along
  semicircular paths,'' \emph{IEEE Antennas and Wireless Propagation Letters},
  vol.~16, pp. 3059--3062, 2017.

\bibitem{MA4}
L.~Zhu, W.~Ma, and R.~Zhang, ``Movable antennas for wireless communication:
  Opportunities and challenges,'' \emph{IEEE Communications Magazine}, pp.
  1--7, 2023.

\bibitem{FA1}
K.~K. Wong, A.~Shojaeifard, K.-F. Tong, and Y.~Zhang, ``Performance limits of
  fluid antenna systems,'' \emph{IEEE Communications Letters}, vol.~24, no.~11,
  pp. 2469--2472, 2020.

\bibitem{AS1}
M.~Gharavi-Alkhansari and A.~Gershman, ``Fast antenna subset selection in mimo
  systems,'' \emph{IEEE Transactions on Signal Processing}, vol.~52, no.~2, pp.
  339--347, 2004.

\bibitem{ars1}
B.~Fuchs, ``Synthesis of sparse arrays with focused or shaped beampattern via
  sequential convex optimizations,'' \emph{IEEE Transactions on Antennas and
  Propagation}, vol.~60, no.~7, pp. 3499--3503, 2012.

\bibitem{ars2}
S.~Yang, B.~Liu, Z.~Hong, and Z.~Zhang, ``Low-complexity sparse array synthesis
  based on off-grid compressive sensing,'' \emph{IEEE Antennas and Wireless
  Propagation Letters}, vol.~21, no.~12, pp. 2322--2326, 2022.

\bibitem{FA2}
K.-K. Wong and K.-F. Tong, ``Fluid antenna multiple access,'' \emph{IEEE
  Transactions on Wireless Communications}, vol.~21, no.~7, pp. 4801--4815,
  2022.

\bibitem{FA3}
H.~{Qin}, W.~{Chen}, Z.~{Li}, Q.~{Wu}, N.~{Cheng}, and F.~{Chen}, ``{Antenna
  Positioning and Beamforming Design for Fluid-Antenna Enabled Multi-user
  Downlink Communications},'' \emph{arXiv e-prints}, p. arXiv:2311.03046, Nov.
  2023.

\bibitem{MA1}
L.~Zhu, W.~Ma, and R.~Zhang, ``Modeling and performance analysis for movable
  antenna enabled wireless communications,'' \emph{IEEE Transactions on
  Wireless Communications}, vol.~23, no.~6, pp. 6234--6250, 2024.

\bibitem{MA2}
L.~Zhu, W.~Ma, B.~Ning, and R.~Zhang, ``Movable-antenna enhanced multiuser
  communication via antenna position optimization,'' \emph{IEEE Transactions on
  Wireless Communications}, vol.~23, no.~7, pp. 7214--7229, 2024.

\bibitem{MA3}
X.~{Pi}, L.~{Zhu}, Z.~{Xiao}, and R.~{Zhang}, ``{Multiuser Communications with
  Movable-Antenna Base Station Via Antenna Position Optimization},''
  \emph{arXiv e-prints}, p. arXiv:2308.05546, Aug. 2023.

\bibitem{FA4}
X.~Lai, T.~Wu, J.~Yao, C.~Pan, M.~Elkashlan, and K.-K. Wong, ``On performance
  of fluid antenna system using maximum ratio combining,'' \emph{IEEE
  Communications Letters}, vol.~28, no.~2, pp. 402--406, 2024.

\bibitem{FPR}
S.~Yang, W.~Lyu, B.~Ning, Z.~Zhang, and C.~Yuen, ``Flexible precoding for
  multi-user movable antenna communications,'' \emph{IEEE Wireless
  Communications Letters}, vol.~13, no.~5, pp. 1404--1408, 2024.

\bibitem{MAISAC1}
Y.~{Xiu}, S.~{Yang}, W.~{Lyu}, P.~{Lep Yeoh}, Y.~{Li}, and Y.~{Ai}, ``{Movable
  Antenna Enabled ISAC Beamforming Design for Low-Altitude Airborne
  Vehicles},'' \emph{arXiv e-prints}, p. arXiv:2409.15923, Sep. 2024.

\bibitem{MAISAC2}
W.~{Lyu}, S.~{Yang}, Y.~{Xiu}, Z.~{Zhang}, C.~{Assi}, and C.~{Yuen},
  ``{Flexible Beamforming for Movable Antenna-Enabled Integrated Sensing and
  Communication},'' \emph{arXiv e-prints}, p. arXiv:2405.10507, May 2024.

\bibitem{MAISAC3}
W.~Lyu, S.~Yang, Y.~Xiu, Z.~Zhang, C.~Assi, and C.~Yuen, ``Movable antenna
  enabled integrated sensing and communication,'' \emph{IEEE Transactions on
  Wireless Communications}, pp. 1--1, 2025.

\bibitem{MAMEC}
Y.~{Xiu}, Y.~{Zhao}, S.~{Yang}, M.~{Xu}, D.~{Niyato}, Y.~{Li}, and N.~{Wei},
  ``{Delay Minimization for Movable Antennas-Enabled Anti-Jamming
  Communications With Mobile Edge Computing},'' \emph{arXiv e-prints}, p.
  arXiv:2409.14418, Sep. 2024.

\bibitem{MAR1}
B.~{Ning}, S.~{Yang}, Y.~{Wu}, P.~{Wang}, W.~{Mei}, C.~{Yuen}, and
  E.~{Bj{\"o}rnson}, ``{Movable Antenna-Enhanced Wireless Communications:
  General Architectures and Implementation Methods},'' \emph{arXiv e-prints},
  p. arXiv:2407.15448, Jul. 2024.

\bibitem{6DMA}
X.~{Shao}, R.~{Zhang}, Q.~{Jiang}, and R.~{Schober}, ``{6D Movable Antenna
  Enhanced Wireless Network Via Discrete Position and Rotation Optimization},''
  \emph{arXiv e-prints}, p. arXiv:2403.17122, Mar. 2024.

\bibitem{MAR2}
S.~{Yang}, J.~{An}, Y.~{Xiu}, W.~{Lyu}, B.~{Ning}, Z.~{Zhang}, M.~{Debbah}, and
  C.~{Yuen}, ``{Flexible Antenna Arrays for Wireless Communications: Modeling
  and Performance Evaluation},'' \emph{arXiv e-prints}, p. arXiv:2407.04944,
  Jul. 2024.

\bibitem{WMMSE1}
Q.~Shi, M.~Razaviyayn, Z.-Q. Luo, and C.~He, ``An iteratively weighted mmse
  approach to distributed sum-utility maximization for a mimo interfering
  broadcast channel,'' \emph{IEEE Transactions on Signal Processing}, vol.~59,
  no.~9, pp. 4331--4340, 2011.

\bibitem{WMMSEFD}
S.~Yang, W.~Lyu, Y.~Xanthos, Z.~Zhang, C.~Assi, and C.~Yuen, ``Reconfigurable
  intelligent surface-aided full-duplex mmwave mimo: Channel estimation,
  passive and hybrid beamforming,'' \emph{IEEE Transactions on Wireless
  Communications}, vol.~23, no.~4, pp. 2575--2590, 2024.

\bibitem{WMMSEISAC}
Y.~Peng, S.~Yang, W.~Lyu, Y.~Li, H.~He, Z.~Zhang, and C.~Assi, ``Mutual
  information-based integrated sensing and communications: A wmmse framework,''
  \emph{IEEE Wireless Communications Letters}, pp. 1--1, 2024.

\bibitem{HWMMSE}
D.~H.~N. Nguyen, L.~B. Le, T.~Le-Ngoc, and R.~W. Heath, ``Hybrid mmse precoding
  and combining designs for mmwave multiuser systems,'' \emph{IEEE Access},
  vol.~5, pp. 19\,167--19\,181, 2017.

\bibitem{DWMMSE1}
Q.~Hu, Y.~Cai, Q.~Shi, K.~Xu, G.~Yu, and Z.~Ding, ``Iterative algorithm induced
  deep-unfolding neural networks: Precoding design for multiuser mimo
  systems,'' \emph{IEEE Transactions on Wireless Communications}, vol.~20,
  no.~2, pp. 1394--1410, 2021.

\bibitem{DWMMSE2}
W.~Jin, J.~Zhang, C.-K. Wen, and S.~Jin, ``Model-driven deep learning for
  hybrid precoding in millimeter wave mu-mimo system,'' \emph{IEEE Transactions
  on Communications}, vol.~71, no.~10, pp. 5862--5876, 2023.

\bibitem{RWMMSE}
X.~Zhao, S.~Lu, Q.~Shi, and Z.-Q. Luo, ``Rethinking wmmse: Can its complexity
  scale linearly with the number of bs antennas?'' \emph{IEEE Transactions on
  Signal Processing}, vol.~71, pp. 433--446, 2023.

\bibitem{newt}
B.~Mamandipoor, D.~Ramasamy, and U.~Madhow, ``Newtonized orthogonal matching
  pursuit: Frequency estimation over the continuum,'' \emph{IEEE Transactions
  on Signal Processing}, vol.~64, no.~19, pp. 5066--5081, 2016.

\bibitem{SAGE}
L.~Weiland, T.~Wiese, and W.~Utschick, ``Multipath mitigation using omp and
  newton's method for multi-antenna gnss receivers,'' in \emph{2017 IEEE 7th
  International Workshop on Computational Advances in Multi-Sensor Adaptive
  Processing (CAMSAP)}, 2017, pp. 1--5.

\bibitem{GP}
T.~Blumensath and M.~E. Davies, ``Gradient pursuits,'' \emph{IEEE Transactions
  on Signal Processing}, vol.~56, no.~6, pp. 2370--2382, 2008.

\end{thebibliography}

\vspace{12pt}

\end{document}